\documentclass[12pt, dvipsnames]{article}

\usepackage{amsfonts}
\usepackage{amssymb}
\usepackage{amsmath}
\usepackage{geometry}
\usepackage[onehalfspacing]{setspace}
\usepackage{float}
\usepackage{supertabular}
\usepackage{rotating}
\usepackage{longtable}
\usepackage{centernot}
\usepackage{graphicx}
\usepackage{caption}
\usepackage{subcaption}
\usepackage{enumerate}
\usepackage{xcolor}
\usepackage{booktabs}
\usepackage[flushleft]{threeparttable}
\usepackage[title]{appendix}
\newtheorem{definition}{Definition}

\newtheorem{proposition}{Proposition}

\usepackage[colorlinks,allcolors=blue]{hyperref}
\usepackage{xcolor}
\usepackage[normalem]{ulem}
\usepackage[round]{natbib}
\usepackage{multirow}
\allowdisplaybreaks[1]

\usepackage{todonotes}

\newcommand{\R}{\mathbb{R}}

\usepackage{graphicx}
\renewenvironment{abstract}
 {\small
  \begin{center}
  \bfseries \abstractname\vspace{-.5em}\vspace{0pt}
  \end{center}
  \list{}{%
    \setlength{\leftmargin}{5mm}
    \setlength{\rightmargin}{\leftmargin}%
  }%
  \item\relax}
 {\endlist}

\title{Stable Marriage, Children, and Intrahousehold Allocations\thanks{We thank Laurens Cherchye, Thomas Demuynck, Bram De Rock, and seminar participants at the RES Annual Conference, the Spring Meeting of Young Economists, University of Antwerp, University of Bristol, University of Essex, University of Leuven, and University of Louvain-la-Neuve for many helpful suggestions.}}
\author{
Mikhail Freer\thanks{Department of Economics, University of Essex. E-mail: \href{m.freer@essex.ac.uk}{m.freer@essex.ac.uk}} \and 
Khushboo Surana\thanks{Department of Economics and Related Studies, University of York. E-mail: \href{khushboo.surana@york.ac.uk}{khushboo.surana@york.ac.uk}}}

\begin{document}
	\sloppy
	\maketitle
	\thispagestyle{empty}

	\begin{abstract}
	\noindent
    We present a revealed preference framework to study sharing of resources in households with  children.
    We explicitly model the impact of the presence of children in the context of stable marriage markets under both potential types of custody arrangement -- joint custody and sole custody. 
    Our models deliver testable revealed preference conditions and allow for the identification of intrahousehold allocation of resources.
    Empirical applications to household data from the Netherlands (joint custody) and Russia (sole custody) show the methods' potential to identify intrahousehold allocation.

    {\bf JEL classifications:} C14, D11, C78 \\  
    {\bf Keywords:} marital stability, children, custody laws, revealed preference analysis, intrahousehold allocation.
    \end{abstract}
	
\newpage
 \setcounter{page}{1}
	
\section{Introduction}

Children are often considered public goods for their parents and play a crucial role in intrahousehold bargaining. 
A key feature of parental divorce is the distribution of property rights over their children.
This distribution is governed by the allocation of custody, which determines both the living arrangement and the decision power parents have after divorce.
We propose a structural model of stable marriage, taking into account the marital-specific nature of children.
We study both forms of post-divorce custody arrangement -- joint custody and sole custody.
An important aspect of our methodology is that it is nonparametric in nature; as such, it does not impose any parametric/functional form assumption on individual utilities or the decision-making process.
Through empirical applications to household data from the Netherlands and Russia, we show that our method can be useful in identifying intrahousehold consumption under both custody arrangements.

Households consist of multiple decision-makers with potentially different preferences.
It is now well established in the literature that unitary models, which assume that households maximize a common household utility function, are often rejected in favour of non-unitary models, which allow for intrahousehold bargaining and preference heterogeneity \citep[see, e.g.,][]{browning1998efficient,attanasio2002tests,cherchye2009opening}.  
We follow a collective approach to household consumption analysis \citep[\`{a} la][]{chiappori1988rational,chiappori1992collective}.  
In the absence of direct information on ``who gets what'' in the household, the collective approach helps in the analysis of intrahousehold allocations of time and resources.
Following \citealp{becker1973theory}, researchers have often integrated the analysis of household consumption patterns with marital matching models.
A combined framework of the two theories of ``who marries whom'' and ``who gets what'' has been empirically used in some recent studies \citep[see, e.g.,][]{cherchye2014household,gousse2017marriage,weber2017collective}.
However, a drawback of such works has been that children and custody laws have received limited attention.
We aim to contribute to this strand of the literature by presenting a structural framework to study household consumption under the assumption of marital stability and the marital-specific public good nature of children.
We consider the implications of the stability of marriages for household consumption under the two most commonly used custody arrangements (joint and sole custody).

Many household consumption models treat children as public consumption within households \citep[see, e.g.,][]{blundell2005collective,cherchye2012married}.
Well being of children, characterized by expenditures on children, is assumed to give utility to both parents.
However, children are different from other types of household public goods (such as housing or transportation) in that they are ``marital-specific'' \citep{becker1991treatise,chiappori2007divorce}.\footnote{
    This assumption has indirect empirical support.
    \citealp{bramlett2002cohabitation} show that children in step-families are equally likely to be adversely affected as children from single-parent families. Children from such households do worse than children living with both biological parents. \citealp{ermisch2008intra} look at remarried fathers with kids from the previous marriage and find that the child support amount transferred for the children is positively correlated with the share of income the father brings to the new couple. Had the preference for the welfare of stepchildren internalized by the new partner, the transfers should not be correlated with the share of income the father brings. Further, this is consistent with the idea of kin selection \citep[introduced by][]{HAMILTON19641, hamilton1964genetical}, a theory from evolutionary biology that can be used to explain unconditional altruism. Kin selection has supporting evidence for humans \citep[see, e.g.,][]{smith1987inheritance,madsen2007kinship} as well as for other biological species.
    } 
Thus, while other public goods can be shared with any potential partner, children bring utility only to the parents of the children and continue to be a public good after parental divorce.
When a couple separates, the living arrangement and the property rights over their children are governed by custody law.
Custody law thus becomes an important factor determining the technology of public-good provision post-divorce.
We consider two custody arrangements (joint and sole custody). 
Under joint custody, both parents share legal and physical custody of the children. In this setting, children spend time with both parents, and both parents have the right to make major decisions about their children. 
Under sole custody, one of the parents (custodian) is awarded legal and physical custody of the children, while the noncustodial parent monetarily supports them by transferring money to the custodian. 
Given that a significant share of a household's budget is spent on children\footnote{
    In 2015, on average, children's consumption accounted for 26\% of the total household expenditure for one-child families to 49\% for three-children families in the United States. \citep[see][]{lino2017usda}.
}, it is vital to incorporate these factors while analyzing household consumption decisions from the perspective of marital stability.

We build on the model of \citealp{cherchye2014household} by incorporating children's consumption and child custody law when characterizing marital stability in terms of intrahousehold consumption behavior. This modification is particularly relevant as ample empirical evidence highlights the effects of custody laws on individuals' decisions \citep[see, e.g.,][]{del1995rationalizing,rasul2005marriage,halla2007bargaining,foerster2020untying}. 
A key feature of our method is that we use a revealed preference framework in the spirit of \citealp{samuelson1938note,afriat1967construction,varian1982nonparametric}, which avoids erroneous conclusions due to parametric misspecification of the utility functions and allows for fully heterogeneous individual preferences. 
Our conditions define testable implications that should hold for every stable matching. Further, as the conditions are linear in unknowns, they can be used for the nonparametric (set) identification of intrahousehold resource shares.


The paper considers a static frictionless marriage market \citep[see, e.g.,][]{legros2007beauty,choo2013collective,chiappori2017matching,galichon2019costly}. Unlike a dynamic model, which would require panel data, the static nature of the model allows us to identify the outcomes of intrahousehold bargaining with as little as a single observation per household (cross-section data).
Admittedly, focusing on static conditions of the marriage market does not allow us to consider important intertemporal decisions such as fertility.\footnote{
    As shown by the previous studies, intertemporal decisions, intrahousehold bargaining and the decision to divorce are affected by child support, alimony, and divorce regimes \citep[see][]{voena2015yours,reynoso2018impact,foerster2020untying}. There are two key differences between these papers that consider life-cycle models and our framework. First, unlike the current paper, these studies do not consider the implications of children as marital-specific public goods within a stable marriage environment. Second, our approach is nonparametric in nature. Hence, there is no risk of functional misspecification. In this light, our contribution complements the parametric methods adopted by these studies.
}
Our approach takes children as given and does not account for any potential fertility plans partners might have. Nonetheless, the equilibrium concept of marital stability that we consider provides a natural starting point from which to analyze individuals' marital and consumption behavior. It can be used as a necessary building block for more advanced dynamic models \citep[for a review, see][]{chiappori2017static}.

We show the potential of our method through empirical applications on household data drawn from the Longitudinal Internet studies for the Social Sciences (LISS) panel and the Russia Longitudinal Monitoring Survey (RLMS). 
We consider a labor supply setting in which households spend their total potential income on spouses' leisure, a Hicksian aggregate private and public good, and a Hicksian aggregate children's consumption.
We show that the revealed preference characterization provides informative bounds on household allocation by exploiting the structural implications of the marriage market under different types of custody arrangements.

The remainder of this paper is organized as follows.
Section \ref{section_setting} introduces the empirical setting and defines the stability of marriages.  
Section \ref{section_rpconditions} presents the revealed preference characterization of stable marriage markets under joint and sole custody arrangements. 
Section \ref{section_empirical} presents the empirical applications.
Section \ref{section_conclusion} provides concluding remarks.
All proofs are collected in the Appendix.


\section{Setting}
\label{section_setting}
We begin by introducing the household consumption setting and a formal definition of stable matching.

\paragraph{Structural Components.} Consider a marriage market with a finite set of males $M$ and a finite set of females $W$.
A matching function $\sigma: M \cup W \rightarrow M \cup W$ describes who is married to whom and satisfies the following properties:

\begin{itemize}
    \item $\sigma(m)\in W$ for every $m\in M$;
    \item $\sigma(w)\in M$ for every $w\in W$;
    \item $w=\sigma(m)$ if and only if $m = \sigma(w)$.
\end{itemize}
In what follows, we focus our attention on married couples ($|M| = |W|$). 
Including singles in the theoretical analysis can be done in a similar way.
We will include singles in our empirical applications to allow for the possibility that a married individual may consider a single of another gender as a potential outside option.

For any couple $(m,w)$, denote by $q_{m,w} \in \R^n_+$ the private consumption, by $Q_{m,w} \in \R^N_+$ the public consumption, and by $C_{m,w}\in \R^L_+$ the children's consumption in the household.
Private consumption is divided between male and female. 
Let $q^m_{m,w}$ and $q^w_{m,w}$ be the private consumption of $m$ and $w$ such that $q^m_{m,w} + q^w_{m,w} = q_{m,w}$. 
Next, denote by $C^m_{m,w}$ the consumption of the biological children of $m$ and by $C^w_{m,w}$ the consumption of the biological children of $w$. 
If all the children in the household are biological to both $m$ and $w$, then $C^m_{m,w} = C^{w}_{m,w} = C_{m,w}$.  
This last restriction stems from our assumption that children are marital-specific public goods.
Given the household's aggregate consumption $(q_{m,w}, Q_{m,w}, C_{m,w})$, household's allocation is defined by $(q^m_{m,w}, q^w_{m,w}, Q_{m,w}, C^m_{m,w}, C^w_{m,w})$.
In the data, we will only observe the consumption bundles of the matched couples (i.e. when $w = \sigma(m)$). 
Hence, we will treat the consumption bundles of all potential matches as unknown quantities with the restriction that they are feasible under the budget set.

Consumption decisions are made under budget constraints formed by prices and income. 
For any couple ($m,w$), let $p_{m,w} \in \R^n_{++}$ be the price vector of the private good and $P_{m,w} \in \R^N_{++}$ be the price vector of the public good. 
Denote by $\rho^{m',\sigma(m)}_{m,w} \in \R^L_{++}$ the price of children's consumption faced by $m$ and $\sigma(m)$ when $m$ forms a couple with $w \in W\cup \{\emptyset\}$ and $\sigma(m)$ forms a couple with $m' \in M \cup \{\emptyset\}$.\footnote{Since we consider a static model, there are no new children realized within the period of observation.}
This notation implies that, even though parents would jointly provide their children's consumption post-divorce, the price they face can depend on their potential matches.  
Finally, let $y_{m,w}$ be the total potential labor and non labor income of the couple $(m,w)$.

We include child custody legislation by specifying the technology used by the parents to produce children's consumption and the child support payment (if any) between partners after divorce.
First, let $\tau^{m',\sigma(m)}_{m,w}$ be the transfer from $m$ to $\sigma(m)$ {when $m$ forms a couple with $w$ and expects his partner $\sigma(m)$ to match with $m'$}.
If $\tau^{m',\sigma(m)}_{m,w}$ is positive (negative), $m$ transfers (receives) money to $\sigma(m)$.
Next, assume that to produce the post-divorce children's consumption, $m$ and $w$ use a technology characterized by cost functions $f^m(C|(m,w), (m',\sigma(m)))$ and $f^{w}(C|(m,w), (\sigma(w),w'))$, respectively. The cost function $f^m$ depends on the potential match of both $m$ and $\sigma(m)$. Similarly, $f^w$ depends on the potential match of both $w$ and $\sigma(w)$. The cost functions and transfers are not observed by the researcher but can be partially inferred using the custody law in place. We will explain this further in Section \ref{section_rpconditions}.
We assume that the cost functions are such that
$$
f^m(C|(m,w), (m', \sigma(m))) + f^{\sigma(m)}(C |  (m', \sigma(m)), (m,w)) = \rho^{m',\sigma(m)}_{m,w} C.
$$
%
The budget constraints for $m$ and $w$ as singles are
\begin{equation*}
\begin{split}
	&	p_{m,\emptyset} q^m + P_{m,\emptyset} Q + f^m(C^m | (m,\emptyset),(m',\sigma(m)) ) \leq y_{m,\emptyset}-\tau^{m',\sigma(m)}_{m,\emptyset}, \\ 
	&	p_{\emptyset, w} q^w + P_{\emptyset,w} Q + f^w(C^w|(\emptyset,w),(\sigma(w),w')) \leq y_{\emptyset,w} +\tau^{\emptyset,w}_{\sigma(w),w'}
\end{split}
\end{equation*}
and the budget constraint for a potential match $(m,w)$ is
\begin{align*}
p_{m,w} (q^m+q^w) + P_{m,w} Q +  &  f^m(C^m|(m,w),(m',\sigma(m))) + f^w(C^w | (m,w),(\sigma(w), w') )  \leq
\\
& \leq y_{m,w} - \tau^{m',\sigma(m)}_{m,w} + \tau^{m,w}_{\sigma(w),w'}.
\end{align*}

%

\noindent
Each individual is assumed to get utility from the consumption of private goods, public goods, and own children's consumption. The preferences of individual $i$ can be described by a continuous, concave, and strictly monotone utility function $u_{i}: \R^n_+\times\R^N_+\times\R^L_+\rightarrow \R$.\footnote{
	We have restricted individual preferences to be egoistic in nature. 
	However, we do allow for the fact that individual utilities depend on public consumption in the household. 
	Public consumption can be either interpreted as pure public consumption in the household or as private consumption with externalities. 
}

\paragraph{Stable Matching.} 
We use the stability of marriage markets as our key identifying assumption for household allocations. 
A matching is called stable if it satisfies individual rationality and has no blocking pairs.\footnote{
	\citealp{cherchye2017stable} include Pareto efficiency as an additional requirement for a stable marriage market. 
	However, they also note that in a cross-sectional setting, Pareto efficiency alone does not generate additional testable implications.  
	That is, the identification of intrahousehold allocation is mainly driven by the requirements of individual rationality and no blocking pairs condition. 
	In this paper, we will also maintain the requirement of Pareto efficiency within every couple (current and potential matches). 
	However, as we assume that children's consumption is always produced by the current spouses (even in the case of divorce or remarriage of one of the spouse), we	drop the Pareto efficiency assumption for this type of consumption.
	That is, when individuals consider outside options, we allow for inefficiency in the public good (children's consumption) to be produced by the ex-partners. 
}
In other words, stability requires that a) all individuals prefer being in the current match rather than staying alone and b) there are no $m\in M$ and $w\in W$ (with $w \neq \sigma(m)$) such that each of them would be better off by marrying each other rather than in the current match. 
More formally, we have the following definitions.

\textit{Individual rationality.} 
A matching $\sigma$ is individually rational if all individuals prefer the allocation obtained in the current matching to any of the allocations the individual can get staying alone.
Formally, individual rationality (for $m\in M$ and $w \in W$) implies,
\begin{align*}
u_{m,\emptyset} &\leq u_m(q^m_{m,\sigma(m)}, Q_{m,\sigma(m)},C^m_{m,\sigma(m)}), \\
u_{\emptyset,w} &\leq u_w(q^w_{\sigma(w),w}, Q_{\sigma(w),w},C^w_{\sigma(w),w}),
\end{align*}
where $u_{m,\emptyset}$ and $u_{\emptyset,w}$ are the maximum achievable utilities of $m$ and $w$ when single. The right-hand side of the two inequalities are the utilities obtained in the current matching.

\textit{No blocking pairs.} 
No blocking pairs condition requires that there is no pair of male and female who find it preferable to marry each other rather than being with their current matches. 
We say that ($m,w$) is a blocking pair if there exists a consumption vector $(q^m_{m,w}, q^w_{m,w}, Q_{m,w}, C^m_{m,w}, C^w_{m,w})$ achievable for the couple $(m,w)$ such that
\begin{align*}
& u_m(q^m_{m,\sigma(m)}, Q_{m,\sigma(m)},C^m_{m,\sigma(m)}) \le u_m(q^m_{m,w},Q_{m,w},C^m_{m,w}) \text{ and,}\\
& u_w(q^w_{\sigma(w),w}, Q_{\sigma(w),w},C^w_{\sigma(w),w}) \le u_w(q^w_{m,w},Q_{m,w},C^w_{m,w}) 
\end{align*}
with at least one inequality being strict. The left hand side of these inequalities are the utilities of $m$ and $w$ in their current matches.

\begin{definition}
A matching $\sigma$ is \textbf{stable} if it is individually rational and there are no blocking pairs.
\end{definition}

\paragraph{Rationalizability.}

Having defined the concept of stable matching, now we discuss the type of dataset that we will study and the meaning of rationalizability of a dataset by stable matching. 
We consider a dataset $\mathcal{D}$ with the following information, 
\begin{itemize}
	\item a matching function $\sigma$,
	\item household consumption ($q_{m,\sigma(m)}, Q_{m,\sigma(m)}, C_{m,\sigma(m)}$) for all matched couples $(m,\sigma(m))$,
	\item prices $(p_{m,w}, P_{m,w}, {\rho_{m,w}^{m',\sigma(m)}, \rho_{m,w}^{\sigma(w),w'}}) $ and income $(y_{m,w})$ faced by all possible pairs $(m,w)$ for $m \in M \cup \lbrace \emptyset \rbrace$ and  $w \in W \cup \lbrace \emptyset \rbrace$.
\end{itemize}
Given a dataset $\mathcal{D}$, we say that it is {\bf rationalizable} by a stable matching if there exists at least one allocation for all matched couples such that the matching $\sigma$ is stable. 
In the next section, we present revealed preference conditions for a stable matching under two types of child custody arrangements.  
These restrictions are necessary conditions that are linear in unknown quantities. 
The linear nature of these conditions makes them easy to be used for the (set) identification of  intrahousehold resource allocations.

\section{Revealed Preference Conditions}
\label{section_rpconditions}
In this section, we lay out two models of stable matching with children's consumption.  
The two models correspond to different types of child custody legislation.
In the first case, we focus on a setting where both parents are custodians and there are no direct transfers between the parents (\textit{joint custody}).
In the second case, we focus on a setting where one of the parents is a custodian and the other parent is obliged to make transfers to the custodian (\textit{sole custody}).   
Finally, we discuss how we can account for deviations from exact rationalizability when taking these models to the data.
Here we also indicate how we can use these models to (set) identify the intrahousehold resource allocation.

\subsection{Joint Custody}
Under joint custody, the rights and responsibilities over children are shared by the spouses and there are no transfers involved.  
Joint custody is an uprising custodian arrangement.  
It has been adopted as a default post-divorce child custody arrangement in Australia, Belgium, France, Germany, the Netherlands and many western countries. 
In this custody setting, children spend time with both parents (joint physical custody). 
Moreover, both parents have equal rights when it comes to the major decisions concerned with children, such as education, health care, etc. (joint legal custody).
In the following, we refer to joint custody as a setting in which both joint physical and legal custody is in place.\footnote{
    Formally, we can also consider a third ``mixed'' model with joint legal custody but sole physical custody. Implications for this mixed setting can be obtained as a combination of the two ``pure'' models we consider.    
}

We formulate the joint custody setting as follows. 
We assume that children's consumption consists of two types of expenditures: daily routine ($k$) and big decisions ($K$).
Upon divorce, expenditures of daily routine would be voluntarily provided by the parent who is responsible for the children at the moment.
Since there is neither monitoring nor enforcement mechanism present in the provision of daily routine consumption, we assume that $k$ is provided non-cooperatively by the parents.
That is,  each parent ($m$ and $\sigma(m)$) makes a voluntary contribution where each unit of  $k$ is bought at the market price.
Next, since both parents decide on children's big decision consumption together, we assume that $K$ would be provided cooperatively.

For the sake of the exposition, consider both $k$ and $K$ to be one-dimensional goods.
We assume (without loss of generality) that the price of $k$ is normalized to one.
For the price of $K$, we assume that it only depends on $m$ and $\sigma(m)$ (parents of the children) and is denoted by $\rho_{m,\sigma(m)}\in \R_{++}$.\footnote{
    This means that $\rho_{m,w}^{m',\sigma(m)} = \rho_{m,\sigma(m)}$ and $\rho_{m,w}^{\sigma(w),w'} = \rho_{\sigma(w),w}$.
    Relaxing this assumption would require us to make ad-hoc assumptions about the beliefs the individuals have about potential matches of their current partners. The stability conditions robust to any belief-related assumptions would require that we use the market price of $K$ instead of the personalized (Lindahl) price of $K$. However, in that case, the cooperative provision of $K$ would be observationally equivalent to the non-cooperative provision. Thus, we prefer to use our assumption, which allows us to exploit the difference in the protocol of the provision of children's consumption. Simplifying the belief structures by restricting these prices to be (current) match specific makes the model tractable. In our empirical application, we show that even with this restricted amount of degrees of freedom on the post-divorce prices, the model may exhibit low identifying power.
}

Under the cooperative provision of $K$, there must exist personalized (Lindahl) prices for each of the parents (ex-partners) such that they add up to the market price (Bowen-Lindahl-Samuelson condition of Pareto efficiency).
Denote these prices by { $\rho^m_{m,\sigma(m)}$ and $\rho^{\sigma(m)}_{m,\sigma(m)}$.}
Finally, as there are no transfers prescribed, $\tau^{m',\sigma(m)}_{m,w} = \tau^{\sigma(w),w'}_{m,w} =0$.
Proposition \ref{prop:JC} shows necessary revealed preference conditions for stability under joint custody.

\begin{proposition}
\label{prop:JC}
If a dataset $\mathcal{D}$ is rationalizable by a stable matching under joint custody, then there exist individual private consumption $q^m_{m,\sigma(m)}, q ^{\sigma(m)}_{m,\sigma(m)}$, personalized prices $\rho^m_{m,\sigma(m)}, \rho^{\sigma(m)}_{m,\sigma(m)}$ for each matched pair $(m,\sigma(m))$ and personalized prices $P^m_{m,w}, P^w_{m,w}$ for each pair ($m,w$) such that
$$
q^m_{m,\sigma(m)} + q^{\sigma(m)}_{m,\sigma(m)} = q_{m,\sigma(m)}; \ \rho^m_{m,\sigma(m)} + \rho^{\sigma(m)}_{m,\sigma(m)} = \rho_{m,\sigma(m)} \text{ and } P^m_{m,w} + P^w_{m,w} = P_{m,w}
$$
that simultaneously meet the following constraints:
\begin{itemize}
	\item [(i)] Individual rationality restrictions for all $m\in M$ and all $w\in W$,
	\begin{equation*}
	\begin{split}
& y_{m,\emptyset} \le p_{m,\emptyset} q^m_{m,\sigma(m)} + P_{m,\emptyset} Q_{m,\sigma(m)}  + k_{m,\sigma(m)} + \rho^m_{m,\sigma(m)} K_{m,\sigma(m)},
\\ 
& y_{\emptyset, w}\le p_{\emptyset,w} q^w_{\sigma(w),w} + P_{\emptyset,w} Q_{\sigma(w),w}  + k_{\sigma(w),w} + \rho^{w}_{\sigma(w),w} K_{\sigma(w),w},
	\end{split}
	\end{equation*}

	\item [(ii)] No blocking pair restrictions for all $(m,w) \in M \times W$,
	\begin{equation*}
	\begin{split}
	y_{m, w} &\le p_{m,w} (q^m_{m,\sigma(m)} + q^w_{\sigma(w),w})  + P^m_{m,w} Q_{m,\sigma(m)} +  P^w_{m,w} Q_{\sigma(w),w} + \\ &
 + k_{m,\sigma(m)} + \rho^m_{m,\sigma(m)} K_{m,\sigma(m)} + k_{\sigma(w),w} + \rho^{w}_{\sigma(w),w} K_{\sigma(w),w}.
 	\end{split} 
 	\end{equation*}
 	\end{itemize}

\end{proposition}

The linear conditions presented in Proposition \ref{prop:JC} have intuitive interpretations.
The first constraint specifies that individual private consumption in the matched couples should add up to the total private consumption in the household. 
The second constraint specifies that the parent's personalized prices for children's consumption should add up to the market price of the good.
Similarly, the third constraint specifies that for all potential pairs, the personalized prices for the public good (e.g. housing or transportation) should add up to the market price of the good. 
These personalized prices quantify an individual's willingness to pay for the public good in the household. Having these prices add up to the market price corresponds to a Pareto optimal provision of public goods among potential pairs.
The right-hand sides of constraints $(i)$ and $(ii)$ specify an upper bound for the costs of the currently consumed bundle in the outside options. 
If the constraints are violated, the individuals can purchase the current consumption bundle with the outside option prices and income, thus making the current match unstable.

\subsection{Sole Custody}
Sole custody requires one of the parents to be in full control of the children and the other parent to support the children's well-being via transfers to the custodial parent. 
Sole custody is currently a dominant post-divorce custody rule in the family laws of Japan, Russia, and several other countries.
In this custody setting, after divorce, children stay with one of the parents (custodian), who has both legal and physical control over the children.
The other parent (non-custodian) is supposed to make transfers to the custodian to support the children. 
The minimum amount of transfer that the non-custodian needs to pay to the custodian is determined by the legal authorities of the state.

To formulate the sole custody setting, we assume that children's consumption can be produced only by the custodian.
The non-custodian makes an unconditional transfer to the custodian.
After receiving the transfer, the custodian maximizes their utility by choosing a consumption bundle that includes the children's consumption.
Further, we assume that the non-custodian knows the preferences of the custodian and chooses a transfer that maximizes his/her own utility.
As children's consumption remains a public good for both parents, this formulation corresponds to a \emph{delegated provision of public good}.
We make a few more additional assumptions.

\begin{enumerate}
	\item All children are assigned to the same custodian;\footnote{
		Alternatively, there can be a `split custody' under which different children would be assigned to different parents.  
		This is essentially a mixture of several sole custody rules.
		Stability conditions for such a setting can be derived in a similar way.
		However, this rule is rarely used in practice. 
		In order to keep the exposition simple, we focus on a setting where all children are assigned to the same custodian.
	}

	\item All custodians are females;\footnote{
		In Russia (which is also the setting of our empirical study), it is still the case that in the majority of child custody cases, the custodian is the mother. The stability conditions without this assumption can be derived in a similar way if we have more precise information on who would be the custodian in every couple.
	}

	\item All custodians are compliant. 
	This means that expenditures on children's consumption by the custodian would be at least as large as the transfer received from the non-custodian;

	\item Transfer received from the non-custodian is no less than $T_{m,\sigma(m)}$, where $T_{m,\sigma(m)}$ is the amount explicitly defined by the child custody guidelines.
\end{enumerate}
Assumptions 1 and 2 can be relaxed if we have more precise guidelines on how the custodian arrangements are determined by the court.
Assumption 4 comes directly from the child custody law. However, we do maintain that there is perfect enforcement of the law.\footnote{
Alternatively, we can allow for imperfect compliance by introducing outside option-specific binary variables indicating whether or not the stability restrictions are satisfied with minimum transfers. These variables can then be identified, conditional on minimizing the divorce costs, by, e.g. maximizing the sum of compliance indicators in the marriage markets. However, we choose not to follow this route in our empirical analysis. This, in principle, also allows for testing the empirical validity of the assumption.
}
We also assume that children's consumption can be described as a one-dimensional Hicksian aggregate of the expenditures on children.
Thus, we can, without loss of generality, assume that the price of $C$ is normalized to one.
Proposition \ref{prop:SPC} shows necessary revealed preference conditions for stability under sole custody.

\begin{proposition}
\label{prop:SPC}
If a dataset $\mathcal{D}$ is rationalizable by a stable matching under sole custody, given the minimum transfers $T_{m,\sigma(m)}$ for every $m\in M$, there exist individual private consumption $q^m_{m,\sigma(m)}, q ^{\sigma(m)}_{m,\sigma(m)}$ for each matched pair $(m,\sigma(m))$ and personalized prices 
$P^m_{m,w}, P^w_{m,w}$ for each pair ($m,w$) such that
$$
q^m_{m,\sigma(m)} + q^{\sigma(m)}_{m,\sigma(m)} = q_{m,\sigma(m)} \text{ and } P^m_{m,w} + P^w_{m,w} = P_{m,w}
$$

that simultaneously meet the following constraints:
\begin{itemize}
	\item [(i)] Individual rationality restrictions for all $m\in M$ and all $w\in W$,
	\begin{equation*}
	\begin{split}
	& y_{m,\emptyset} \le p_{m,\emptyset} q^m_{m,\sigma(m)} + P_{m,\emptyset} Q_{m,\sigma(m)}  + C_{m,\sigma(m)},
\\ 
	& y_{\emptyset, w} + T_{\sigma(w),w} \le p_{\emptyset,w} q^w_{\sigma(w),w} + P_{\emptyset,w} Q_{\sigma(w),w}  + C_{\sigma(w),w},
	\end{split}
	\end{equation*}

	\item [(ii)] No blocking pair restrictions for all $(m,w) \in M \times W$, 
	$$
	y_{m, w} + T_{\sigma(w),w} \le p_{m,w} (q^m_{m,\sigma(m)} + q^w_{\sigma(w),w})  + P^m_{m,w} Q_{m,\sigma(m)}  + P^w_{m,w} Q_{\sigma(w),w} + C_{m,\sigma(m)} +  C_{\sigma(w),w}.
 	$$
\end{itemize}
\end{proposition}

\subsection{Stability Indices and Identification}
\label{section_stabilityindices}
\paragraph{Stability Indices.}
The rationalizability conditions shown in Propositions \ref{prop:JC} and \ref{prop:SPC} are strict in nature. The observed household data would either satisfy the constraints or fail to find a feasible solution. 
In reality, household consumption behavior may not be exactly consistent with the model if, for example, the data contain measurement errors, there are frictions in the marriage market, or other factors, such as match quality, affect marital behavior.
In such cases, it is useful to quantify the deviations of observed data from exactly rationalizable behavior.
Following \citealp{cherchye2017stable}, we evaluate the goodness-of-fit of a model by using stability indices. These indices allow us to quantify the degree to which the observed behavior is consistent with the exactly rationalizable behavior.

Formally, we include a stability index $s_{m,w}$ in the inequality constraint for the corresponding outside option $(m,w) \in \{ M \cup \emptyset \} \times \{ W \cup \emptyset \}$. 
In particular, we replace inequalities {\it (i)} and  {\it (ii)} in Proposition \ref{prop:JC} by:
	\begin{equation*}
	\begin{split}
 	s_{m,\emptyset} \times y_{m,\emptyset} & \le p_{m,\emptyset} q^m_{m,\sigma(m)} + P_{m,\emptyset} Q_{m,\sigma(m)}  + k_{m,\sigma(m)} + \rho^m_{m,\sigma(m)} K_{m,\sigma(m)},
 	\\ 
 	s_{\emptyset,w} \times y_{\emptyset, w} & \le p_{\emptyset,w} q^w_{\sigma(w),w} + P_{\emptyset,w} Q_{\sigma(w),w}  + k_{\sigma(w),w} + \rho^{w}_{\sigma(w),w} K_{\sigma(w),w},
 	\\
 	s_{m,w} \times y_{m, w} & \le p_{m,w} (q^m_{m,\sigma(m)} + q^w_{\sigma(w),w})  + P^m_{m,w} Q_{m,\sigma(m)} +  P^w_{m,w} Q_{\sigma(w),m}
 	\\ & \quad + k_{m,\sigma(m)} + \rho^m_{m,\sigma(m)} K_{m,\sigma(m)} + k_{\sigma(w),w} + \rho^w_{\sigma(w),w} K_{\sigma(w),w},
 	\end{split} 
 	\end{equation*}
and inequalities {\it (i)} and  {\it (ii)} in Proposition \ref{prop:SPC} by:
 	\begin{equation*}
	\begin{split}
	s_{m,\emptyset} \times y_{m,\emptyset} & \le p_{m,\emptyset} q^m_{m,\sigma(m)} + P_{m,\emptyset} Q_{m,\sigma(m)}  + C_{m,\sigma(m)}, 
	\\ 
	s_{\emptyset,w} \times y_{\emptyset, w} + T_{\sigma(w),w} & \le p_{\emptyset,w} q^w_{\sigma(w),w} + P_{\emptyset,w} Q_{\sigma(w),w}  + C_{\sigma(w),w}, 
	\\
	s_{m,w} \times y_{m, w} + T_{\sigma(w),w} & \le p_{m,w} (q^m_{m,\sigma(m)} + q^w_{\sigma(w),w})  + P^m_{m,w} Q_{m,\sigma(m)}  + P^w_{m,w} Q_{\sigma(w),m}
	\\
	& \quad + C_{m,\sigma(m)} +  C_{\sigma(w),w}.
	\end{split}
	\end{equation*}
Further, we add the restriction $0 \leq s_{m,w} \leq 1$. Imposing $s_{m,w} = 1$ results in the original rationalizability restrictions, while imposing $s_{m,w}= 0$ would rationalize any behavior. 
Intuitively, the stability indices measure the loss of post-divorce income needed to represent the observed marriages as stable.
Generally, a lower stability index corresponds to a higher income loss associated with a particular outside option. 
This can be interpreted as a greater violation of the underlying model assumptions. 

In our empirical applications, we will use stability indices to evaluate the performance of our models. To identify the values of stability indices for a given data set $\mathcal{D}$, we compute
\begin{equation*}
    \max \sum_{m} \sum_{w} s_{m,w}
\end{equation*}
subject to the rationalizability conditions. This gives a stability index $s_{m,w}$ for each exit option.
If the original constraints are satisfied, then there is no need for adjustment, and our stability indices will be equal to one. 
Otherwise, a strictly smaller index will be required to rationalize the behavior.

\paragraph{Set Identification.}
Using the stability indices defined above, we can construct a new data set that is consistent with the rationalizability restrictions. More specifically, we can use the identified values of $s_{m,w}$ to rescale the original income levels $y_{m,w}$ such that the adjusted dataset is consistent with the rationalizability restrictions. 
After that, we can use this new data to identify the unobserved parameters of household allocation (such as individual private consumption or resource shares).

In our empirical applications, we will identify a female's share of private consumption and sharing rule, which measures the share of household resources allocated to her. 
Sharing rule governs how resources are distributed within households. 
We will obtain set identification by computing the lower and upper bounds on these quantities subject to the rationalizability
restrictions. 
For example, a lower bound for $w$'s private consumption share can be obtained by minimizing $q^w$ subject to the stability restrictions. 
This automatically defines an upper bound for the private consumption of $\sigma(w)$ through the adding up constraint ($q^w_{\sigma(w),w} + q^{\sigma(w)}_{\sigma(w),w} = q_{\sigma(w),w}$). 
Effectively, these bounds define feasible values of individual private consumption shares that are compatible with our rationalizability restrictions.

\section{Empirical Illustration}
\label{section_empirical}
We show the usefulness of our method for the identification of intrahousehold allocation through applications on household data drawn from the LISS panel and RLMS.
We consider a labor supply setting where households allocate their entire potential income on both spouses' leisure, Hicksian aggregate private and public goods, and children's consumption. 
In what follows, we present an application of the joint custody model on data from the Netherlands and sole custody model on data from Russia.
We show that both the models have identifying power and lead to reasonably tight bounds on household resource allocation. 

\subsection{Data and Setup}

\paragraph{LISS.} The LISS panel is gathered by CentERdata and is a representative survey of households in the Netherlands. The survey collects rich data on economics and sociodemographic variables at both individual and household levels. For this application, we use the sample of households analyzed by \citealp{cherchye2014household}. The sample selection criteria consider individuals with or without children, working at least 10 hours per week in the labor market, and aged between 25 and 65. The sample contains 832 individuals comprising of 239 couples, 166 single males, and 188 single females. Table \ref{table_sumstat_LISS} in Appendix \ref{appendix_empirics} provides summary statistics for all couples under consideration.

\paragraph{RLMS.}
RLMS is a nationally representative survey of individuals and families in the Russian Federation. 
The survey collects detailed information at both individual and household levels on a variety of topics like health and dietary intake, expenditures, education, income and wages, marriage, fertility, etc. The sample of households used in this study is drawn from the 2013 wave of the survey.\footnote{
    We have chosen the 2013 wave as the last relatively normal year of the Russian economy, which has started going over significant structural changes in 2014. 
    Due to the government's restriction on food imports, the structure of consumption has changed significantly. 
    In addition, the continuing economic decline caused a significant growth of the poverty rate and an increasing share of economies to move to the ``grey'' area.
} 
We use the following selection rules to prepare the final sample.
First, we exclude households with members other than the couple (or single adult) and their (biological) children. 
Next, we confine our study to households whose adults are between 25 and 65 years old and working at least 10 hours per week in the labor market.
We also drop households with missing information (e.g. wage, time use, expenditure) and the outliers by trimming the households in the 1st and 99th percentiles of wages and non-labor income distribution. 
Finally, we exclude households living in Krasnodar, Kushevkiy Rayon, Krasnodarkij Kraj, Georgievskij Rayon, Stavropolskij Kraj, Zolskij Rajon or Kabardino-Balkarija.\footnote{
We remove these regions from the analysis because they are systematically different (e.g. in the legal or economic practices) as compared to the rest of Russia. Appendix \ref{appendix_empirics} provides a more detailed motivation for this sample selection criterion. 
} 
This selection procedure results in a sample of 2492 individuals consisting of 882 couples, 147 single males, and 581 single females. Table \ref{table_sumstat_RLMS} in Appendix \ref{appendix_empirics} presents summary statistics for all couples under consideration. 

\paragraph{Empirical Setting.}
In both datasets, we only observe aggregate household expenditures. As is typical with most household data, we do not have more granular information on the share of public versus private consumption or expenditures on children. In order to deal with this data issue, we adopt the following strategy. We use all expenditure information to form a Hicksian good with price normalized to one. Next, we estimate children's expenditure using the OECD-modified adult equivalence scale  \citep[see][]{donni2015measuring}. 
These estimates are presented in Table \ref{table_adult_equivalence_scale}. 
\begin{table}[htbp]
\centering
\caption{Cost of children as a percentage of total expenditure (OECD-modified scale)}
\label{table_adult_equivalence_scale}
\begin{tabular}{ccc} \\ \toprule 
Number of children & couples & singles \\ \midrule
1 & 17\% & 23\% \\
2 & 28\% & 37\% \\
3 & 37\% & 47\% \\ \bottomrule
\end{tabular}
\end{table}
For the rest, we closely follow the assumptions of \citealp{cherchye2014household}. First, we assume that half of the total adult expenditure is used for public consumption and the other half is for private consumption. Leisure is privately consumed. 
To evaluate individuals' outside options, we need prices and income for all counterfactual situations (i.e., becoming single or remarrying). 
For leisure, we take the price to be individual wages.\footnote{
    Here we assume that the labor market productivity is independent of an individual's marital status. Although this assumption may look rather restrictive in light of the literature on marriage premiums and penalties, with additional modeling assumptions, prices in counterfactual situations can also be imputed.
}
For all other consumption categories, prices are normalized to one. 

Next, for all observed and unobserved matching pairs, household income is defined as the sum of potential labor income (which is 112 times the individual wage rate) and non-labor income of the individuals. For observed couples, we use a consumption-based measure of total non-labor income. More precisely, a household's non-labor income is defined as total consumption expenditures minus total potential labor income. We treat individual non-labor income as unknowns that are subject to the constraint that they must add up to the household non-labor income. Further, rather than fixing individual non-labor to be 50\% of household non-labor income, we allow these unknowns to be endogenously defined by marriage market implications. However, to avoid unrealistic scenarios, we will restrict individual non-labor income to be between 40\% and 60\% of household non-labor income. The same rules were adopted by \citealp{cherchye2017stable}.

To use our models empirically, we require prior specifications of marriage markets and individual consideration sets that specify who the individual would consider as an outside option. 
For our first application (based on the data from the Netherlands), we consider a single marriage market. 
For the second application (based on the data from Russia), we divide our sample into eight marriage markets based on the region of residence. These regions and corresponding sub-sample sizes are shown in Figure \ref{figure_russia} in Appendix \ref{appendix_empirics}.
There is quite some heterogeneity in the size of the marriage market ranging from 55 to 298. 
Considering such region-based sub-samples allows for the possibility of geographically restricted marriage markets.
Next, within each marriage market, we construct individual-specific consideration sets based on age.
In particular, we assume that a male (female) individual's consideration set contains all single and married females (males) who are within a certain age range with him. 
These age restrictions correspond to the 1st and 99th percentile of the distribution of age difference of the matched couples in the sample. 
Appendix \ref{appendix_empirics} provides more information on each market and the average size of individual-specific consideration sets within each market.

\paragraph{Legal Setting.}

The process of divorce in both the Netherlands and Russia is relatively straightforward. 
According to Dutch legislation, a claim of irretrievable breakdown of the marriage by any one spouse is enough to pronounce a divorce. 
If both spouses agree and have no children, divorce can be conducted as an administrative procedure without the necessity to show up at the court.
However, if a couple has children, the parents are obliged to submit a parenting plan when filing for the divorce petition.  
With regards to children's custody, the starting principle is that there is joint custody. 
If the parents are legally married, they automatically have joint custody of their children. If the parents are not legally married, the mother has sole custody, but the father can apply for joint custody in court. If the father makes an application, this will, in principle, be granted.
We model the Dutch household behavior using our characterization of a stable marriage market under joint custody.

According to Russian legislation, divorce can be obtained unilaterally without a need to specify the reason for divorce. 
If both spouses agree and have no children, divorce can be conducted as an administrative procedure without any need to visit the court.
However, if a couple has children, they would be obliged to go through court. 
With regards to children's custody, the Russian family code prescribes assigning custody in the ``best interest'' of the child.
In practice, however, the custody right in most cases is assigned to the mother.\footnote{
        There is no notion of joint custody in the Russian family code.
        Although parents can arrange joint custody, it would require creating a contract within the dimension of Civil Law, not Family Law.}
Further, the non-custodial parent must pay child support once the divorce is confirmed. 
Child support is paid until the children become adults and is straightforward to compute. 
In particular, it is based on a tiered system of 25\% of non-custodian's labor income for the first child, 33\% for two children, or 50\% for three or more children. We account for this by increasing (decreasing) the  custodian's (non-custodian's) post-divorce income by the corresponding monetary value.
We model the Russian household behavior using our characterization of a stable marriage market under sole custody.

\subsection{Rationalizability}
\label{empirics_RLMS}

As a first step, we evaluate the goodness-of-fit by examining the extent to which the observed behavior satisfies the rationalizability conditions under joint custody (Proposition \ref{prop:JC}) for the Dutch households and under sole custody (Proposition \ref{prop:SPC}) for the Russian households.  
We summarize the stability indices by defining two measures for each couple: {\it average} and {\it minimum} stability index. 
The average stability index measures a couple's average income loss that is required to make the marriage stable and the minimum stability index refers to the highest income loss that is required to rationalize all possible exit options. It is the stability index associated with the most attractive outside option of either spouse.

Table \ref{table_dc} summarizes the average and minimum stability indices for the joint custody model in Panel A and the sole custody model in Panel B.
Both the average and minimum stability indices suggest that the models fit the observed household behavior well. The mean values of the average stability indices for the joint and sole custody models are 99.39\% and 99.50\%, respectively.
The average stability indices for both models are very close to 1, implying that the data are very close to satisfying the exact rationalizability conditions.
The minimum stability indices, which reflect the deviation from exact rationalizability with respect to the most attractive outside option, are also generally high. Some couples do seem to need a large adjustment in the post-divorce income to rationalize their current marriage. The lowest minimum stability index is 81.82\% for the joint custody model and 53.81\% for the sole custody model.
Nonetheless, the overall results suggest that we need fairly small adjustments in the data to obtain consistency with the revealed preference conditions.

\begin{table}[htbp]\centering
\caption{Stability indices (in \%) \label{table_dc}}
\begin{tabular}{l*{1}{cc}}
\\ \toprule
\multicolumn{2}{l}{{\it Panel A: joint custody}} & \\
            &     average&     minimum\\
            \cmidrule(lr){2-3} 
mean        &       99.39&       92.04\\
std. dev.          &        0.62&        3.17\\
min         &       93.98&       81.82\\
p25         &       99.19&       89.28\\
p50         &       99.58&       92.19\\
p75         &       99.78&       94.73\\
max         &      100.00&       99.72 \\
\midrule
\multicolumn{2}{l}{{\it Panel B: sole custody}} & \\
            &     average&     minimum\\
            \cmidrule(lr){2-3} 
mean        &       99.50&       88.63\\
std. dev.          &        0.50&        7.04\\
min         &       96.19&       53.81\\
p25         &       99.30&       85.66\\
p50         &       99.64&       89.38\\
p75         &       99.83&       92.68\\
max         &      100.00&      100.00\\
\bottomrule
\end{tabular}
\end{table}

\subsection{Identification}

Next, we use the identified values of stability indices to adjust the data such that they are consistent with the rationality conditions. We use the adjusted data to set identify the female sharing rule defining intrahousehold allocation. To assess the identifying power of our revealed preference methodology, we compare the bounds identified through stable matching restrictions with naive bounds. The naive bounds are obtained without using any
restrictions associated with the stability requirement. They are defined as follows. The naive lower bound for a female equals the share of her private assignable consumption in the household's total income. The corresponding upper bound equals the share of her private assignable and all nonassignable consumption (including children's consumption) in the household's income.   
This corresponds to the scenario where all household consumption except the male's private assignable consumption is allocated to the female.
  

Table \ref{table_bounds_shares} presents the identified values of the women's sharing rule. 
The estimates obtained through our revealed preference characterization are labeled ``stable bounds'', and the ones obtained without any modeling restrictions are labeled ``naive bounds''. 
The columns ``lower'' (``upper'') provide the lower (upper) bounds and columns ``difference'' show the difference between the upper and lower bounds.
They reflect the identifying power of the models under consideration. 
The estimates for the Dutch households (obtained using the joint custody model) are shown in Panel A and those for the Russian households (obtained using the sole custody model) are shown in Panel B.

\begin{table}[htbp]\centering
\caption{Sharing rule  \label{table_bounds_shares}}
\begin{tabular}{l*{1}{cccccc}}
\\ \toprule
\multicolumn{7}{l}{{\it Panel A: joint custody}} \\
& \multicolumn{3}{c}{stable bounds} & \multicolumn{3}{c}{naive bounds} \\
                    &       lower&       upper&  difference&       lower&       upper&  difference\\
                    \cmidrule(lr){2-4} \cmidrule(lr){5-7}
mean                &        0.41&        0.59&        0.18&        0.40&        0.62&        0.22\\
std. dev.                  &        0.08&        0.08&        0.06&        0.08&        0.08&        0.06\\
min                 &        0.12&        0.33&        0.06&        0.12&        0.34&        0.09\\
p25                 &        0.35&        0.54&        0.14&        0.35&        0.57&        0.17\\
p50                 &        0.41&        0.59&        0.18&        0.40&        0.62&        0.22\\
p75                 &        0.46&        0.64&        0.21&        0.45&        0.67&        0.26\\
max                 &        0.66&        0.79&        0.38&        0.65&        0.84&        0.44\\
\midrule 
\multicolumn{7}{l}{{\it Panel B: sole custody}} \\
 & \multicolumn{3}{c}{stable bounds} & \multicolumn{3}{c}{naive bounds} \\
                    &       lower&       upper&  difference&       lower&       upper&  difference\\
                    \cmidrule(lr){2-4} \cmidrule(lr){5-7}
mean                &        0.27&        0.57&        0.30&        0.24&        0.70&        0.46\\
std. dev.                  &        0.12&        0.13&        0.14&        0.11&        0.13&        0.16\\
min                 &        0.01&        0.18&        0.03&        0.01&        0.20&        0.03\\
p25                 &        0.18&        0.48&        0.20&        0.16&        0.61&        0.34\\
p50                 &        0.26&        0.57&        0.27&        0.23&        0.70&        0.45\\
p75                 &        0.34&        0.65&        0.37&        0.31&        0.79&        0.57\\
max                 &        0.74&        0.93&        0.86&        0.70&        1.00&        0.91\\
\bottomrule
\end{tabular}
\end{table}

We find that our revealed preference methodology provides improvements over naive bounds. For the Dutch households, the mean difference between the upper and lower naive bounds is 22 p.p., while the mean difference between the bounds obtained from our methodology is 18 p.p.. While this improvement may look modest, we remark that, in this setup, any improvement in sharing rule identification is implicitly driven by the identification of female's private consumption.\footnote{
    The same argument has been stressed by \citealp{cherchye2014household}. For more details on identification in stable marriage models see \citealp{freer2021stable}.
    } 
In what follows, we show that the identified stable bounds on female private consumption shares are significantly tighter than the corresponding naive bounds.
Next, the results in Panel B show that the stable bounds obtained through the sole custody model also provide substantially tighter identification as compared to the naive bounds. 
The mean difference between the upper and lower naive bounds is 46 p.p., while this difference is much smaller, 30 p.p., for the stable bounds. 
Once again, the substantial improvement in the tightness of the sharing rule identification is implicitly driven by tighter bounds on private consumption shares.

Next, we identify females' private consumption shares by computing lower and upper bounds on $q^w$ subject to the stability restrictions.  Once again, we compare the bounds identified through the stability restrictions with the naive bounds, which are  obtained directly from the data. 
The identified bounds are reported in Table \ref{table_bounds_qf}.
The LISS panel collects assignable information for the Hicksian aggregate private consumption. This information forms the naive bounds in Panel A. RLMS does not provide any information on individual private consumption. Thus, the naive bounds in Panel B correspond to the interval [0,1].  
The results show that the identified stable bounds on female private consumption shares are significantly tighter than the corresponding naive bounds for both models. For the joint custody model, the average difference between the upper and lower naive bounds is 44 p.p., while it is much smaller, 11 p.p. for the stable bounds.
For the sole custody model, the difference between the upper and lower naive bounds is 100 p.p., while it is only 17 p.p. for the stable bounds.
These illustrative applications show the usefulness of our methodology for an informative analysis of intrahousehold allocation.  

\begin{table}[htbp]\centering
\caption{Bounds on females' private consumption \label{table_bounds_qf}}
\begin{tabular}{l*{1}{cccccc}}
\\ \toprule
\multicolumn{7}{l}{{\it Panel A: joint custody}} \\
& \multicolumn{3}{c}{stable bounds} & \multicolumn{3}{c}{naive bounds} \\
                    &       lower&       upper&  difference&       lower&       upper&  difference\\
                    \cmidrule(lr){2-4} \cmidrule(lr){5-7}
mean                &        0.36&        0.47&        0.11&        0.29&        0.73&        0.44\\
std. dev.                  &        0.17&        0.19&        0.14&        0.13&        0.12&        0.18\\
min                 &        0.02&        0.09&        0.00&        0.02&        0.25&        0.03\\
p25                 &        0.24&        0.32&        0.00&        0.20&        0.67&        0.30\\
p50                 &        0.34&        0.46&        0.05&        0.27&        0.75&        0.44\\
p75                 &        0.44&        0.62&        0.18&        0.39&        0.82&        0.57\\
max                 &        0.88&        0.91&        0.58&        0.72&        0.97&        0.82\\
\midrule 
\multicolumn{7}{l}{{\it Panel B: sole custody}} \\
 & \multicolumn{3}{c}{stable bounds} & \multicolumn{3}{c}{naive bounds} \\
                    &       lower&       upper&  difference&     lower&       upper&  difference \\
                    \cmidrule(lr){2-4}  \cmidrule(lr){5-7}
mean                &        0.14&        0.31&        0.17 & 0.00 & 1.00 & 1.00\\
std. dev.                  &        0.19&        0.27&        0.18 & 0.00 & 1.00 & 1.00\\
min                 &        0.00&       0.00&       0.00 &0.00 & 1.00 & 1.00\\
p25                 &        0.00&        0.07&        0.02 & 0.00 & 1.00 & 1.00\\
p50                 &        0.05&        0.28&        0.14 & 0.00 & 1.00 & 1.00\\
p75                 &        0.23&        0.49&        0.25 & 0.00 & 1.00 & 1.00\\
max                 &        1.00&        1.00&        1.00 & 0.00 & 1.00 & 1.00\\
\bottomrule
\end{tabular}
\end{table}

\begin{figure}[htbp]
\centering
\caption{Female sharing rule and wage ratio}
\label{figure_shares}
\begin{subfigure}{.5\textwidth}
  \centering
  \caption{Joint custody}
  \label{fig:sub1}
  \includegraphics[width=\linewidth]{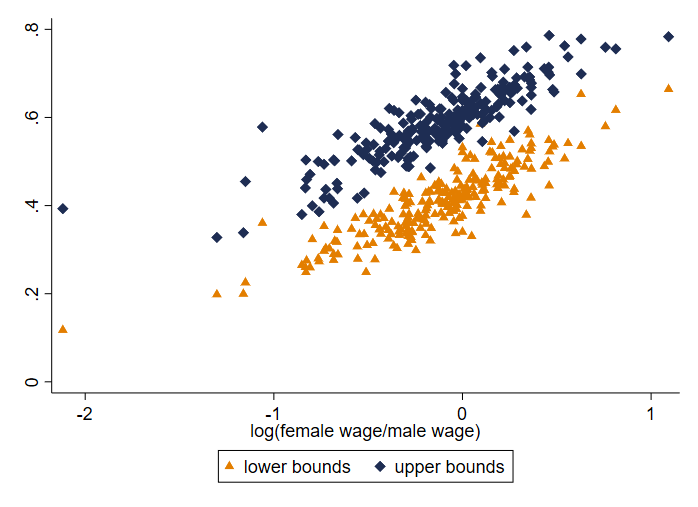}
\end{subfigure}%
\begin{subfigure}{.5\textwidth}
  \centering
  \caption{Sole custody}
  \label{fig:sub2}
  \includegraphics[width=\linewidth]{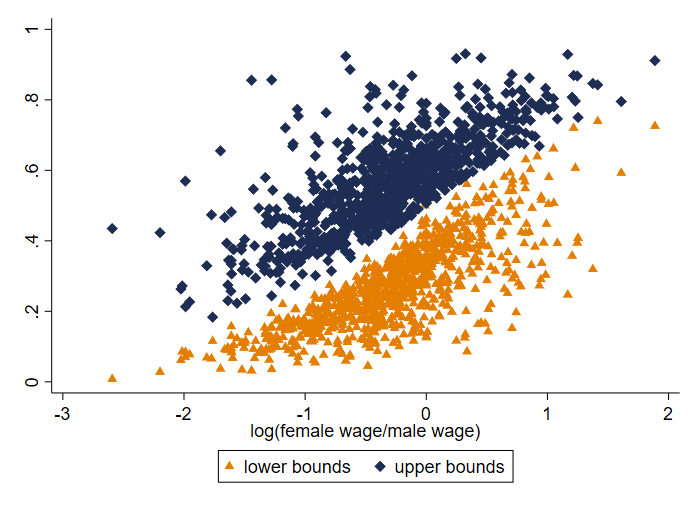}
\end{subfigure}
\end{figure}

Given that there is quite some heterogeneity in women's sharing rule across households, as a follow-up exercise, we  investigate the relationship between the identified female resource shares and observed household characteristics. 
Tables \ref{table_regression_qf_LISS} and \ref{table_regression_qf_RLMS} in Appendix \ref{appendix_empirics} report the OLS regression estimates where we relate the lower and upper bound estimates of female's sharing rule to observed characteristics. The estimates from both models suggest that a woman's resource share increases with the intrahousehold wage ratio (logarithm of female wage divided by male wage). This is visually depicted in Figure \ref{figure_shares}, which plots the estimated stable upper and lower bound for each household in the sample (y-axis) against the intrahousehold wage ratio (x-axis). The patterns suggest that a woman with a higher relative income obtains a relatively larger share of household resources. This can be intuitively understood by considering the marriage market implications. A higher relative wage implies better outside options for the female, which gives her a higher bargaining power within the household.  

Next, the regression estimates for the sole custody model indicate that the female sharing rule increases with the number of children in the household. A possible explanation for this could be the transfers mothers receive as custodians from fathers, which increase the value of their outside options, resulting in a better bargaining position within their current marriages. By contrast, there are no such effects in the joint custody model. As children's responsibility is equally shared between the parents and no transfers are involved, we do not find any significant shift in the female sharing rule by the number of children. These results indicate the empirical usefulness of using marriage market implications for analyzing household consumption.

\section{Conclusion}
\label{section_conclusion}
We developed revealed preference conditions to analyze the implications of stable marriage markets on household consumption decisions while accounting for children's consumption and child custody law. 
In particular, we presented testable revealed preference restrictions under two types of child custody arrangements - joint custody and sole custody.
These conditions allow for the identification of intrahousehold allocation.
To illustrate the empirical performance of these conditions, we provided applications to household data from the Netherlands and Russia.
These applications confirm that both sets of rationality conditions (joint and sole custody) are useful in the identification of household consumption. We showed that our revealed preference conditions provide informative bounds on the female sharing rule, which reflects the share of household resources enjoyed by the female.

In our model, we made several simplifying choices. Weakening these assumptions can enrich the model and create a more realistic setting. We can mitigate some of the assumptions by integrating our model with existing revealed preference methods.
For example, detailed time-use studies often find that couples with children spend a substantial part of their day in joint child care \citealp{cosaert2023togetherness}. While we abstracted from household production, it can be included using the framework of \citealp{cherchye2021did}. Next, we assumed that the nature of consumption (public or private) is known. To additionally identify the degree of publicness of household consumption, we can adapt
the method of \citealp{cherchye2020marital}. Finally, we assumed that the researcher knows individuals' marriage markets. In the empirical application, we operationalized this by defining the region and age-based sets of potential partners. Integrating the insights from the literature that provides a structural explanation of observed matching patterns into the model would allow for more advanced modeling of individual marriage markets \citep[see, e.g.][]{choo2006marries, dupuy2014personality}.

\bibliographystyle{plainnat}
\bibliography{references.bib}

\appendix

\clearpage
\section{Proofs}
\subsection{Proof of Proposition \ref{prop:JC}}
Consider a couple ($m, \sigma(m)$). Upon divorce and under joint custody, children's consumption consists of daily expenditures ($k$), which are provided non-cooperatively through voluntary contributions, and big-decision expenditures ($K$), which are provided cooperatively by the parents ($m,\sigma(m)$). 
We can rewrite individual $i$'s utility function as $u_i(q^i, Q,k^i, K^i)$ for every $i\in M\cup W$, where the superscript $i$ in $k^i$ and $K^i$ indicates that the individual derives utility from own children's consumption. For the sake of the exposition and without loss of generality, we assume that both types of children's consumption are one-dimensional goods.
Without loss of generality, we can normalize the price of $k^i$ to be one.
Price of $K^i$ is denoted by $\rho_{m,w}$ for every $m\in M$ and $w=\sigma(m)\in W$.\footnote{
    The same arguments will hold for an arbitrary finite dimensionality of $k^i$ and $K^i$.}
As children's daily expenditure $k$ is provided non-cooperatively by the parents, we have $k^i = k^i_i + k^i_{\sigma(i)}$ for every $i\in M\cup W$, where $k^i_i$ and $k^i_{\sigma(i)}$ are the contributions of $i$ and $\sigma(i)$ respectively.
Let $k_{i,\sigma(i)}$ be the total daily expenditure observed within the current match $(i,\sigma(i))\in M\times W$.
For the big-decision expenditures ($K$), we assume that the ex-partners would provide this cooperatively. This implies that the sum of their individual willingness to pay for the big-decision expenditure would be equal to the market price. Let us denote by $\rho^m_{m,\sigma(m)}$ and $\rho^w_{\sigma(w),w}$ the willingness to pay for this public good by $m$ and $\sigma(m)$ respectively. Pareto efficiency implies that $\rho^m_{m,\sigma(m)}+\rho^{\sigma(m)}_{m,\sigma(m)} = \rho_{m,\sigma(m)}$.

\paragraph{Individual Rationality.}
We start by examining the individual rationality condition. 
Here we only show the derivation for $m\in M$, the corresponding condition for $w$ can be obtained in a similar way.
The optimization problem of $m$, when single, looks as follows:
\begin{equation*}
\begin{cases}
& u_m(q^m, Q, k^m_{m}+k^m_{\sigma(m)},K^m) \rightarrow \max_{q^m,Q,k^m_m, K^m} \\ 
& p_{m,\emptyset}q^m +P_{m,\emptyset}Q +  k^m_m + \rho^m_{m,\sigma(m)} K^m \leq y_{m,\emptyset} \\
\end{cases}
\end{equation*}
%
%
%
First order conditions give,
\begin{equation*}
\begin{split}
& \nabla_{q^m} u_m(q^m_{m,\emptyset},Q_{m,\emptyset},k_{m,\emptyset}^m+k^{m}_{m',\sigma(m)},K^m_{m,\emptyset}) = \lambda_{m,\emptyset} p_{m,\emptyset}, \\
& \nabla_{Q} u_m(q^m_{m,\emptyset},Q_{m,\emptyset},k_{m,\emptyset}^m+k^{m}_{m',\sigma(m)},K^m_{m,\emptyset}) = \lambda_{m,\emptyset} P_{m,\emptyset}, \\
& \nabla_{k^m} u_m(q^m_{m,\emptyset},Q_{m,\emptyset},k_{m,\emptyset}^m+k^{m}_{m',\sigma(m)},K^m_{m,\emptyset}) =\lambda_{m,\emptyset}, \\
& \nabla_{K^m} u_m(q^m_{m,\emptyset},Q_{m,\emptyset},k_{m,\emptyset}^m+k^{m}_{m',\sigma(m)},K^m_{m,\emptyset}) =\lambda_{m,\emptyset} \rho^m_{m,\sigma(m)},
\end{split}
\end{equation*}
where $k_{m,\emptyset}^m$ is the contribution to own children's daily expenditure by $m$ when single and $k^{m}_{m',\sigma(m)}$ is the contribution to own children's consumption by $\sigma(m)$ when matched with $m'$. Concavity of the utility function implies, 
\begin{equation*}
\begin{split}
& u_m(q^m_{m,\sigma(m)}, Q_{m,\sigma(m)}, k^m_{m,\sigma(m)}, K^m_{m,\sigma(m)}) 
- u_m(q_{m,\emptyset}, Q_{m,\emptyset}, k^m_{m,\emptyset} + k^m_{m',\sigma(m)}, K^m_{m,\emptyset})
\\ 
& \leq \nabla_{q^m} u_m (q^m_{m,\sigma(m)}-q^m_{m,\emptyset}) + \nabla_{Q} u_m (Q_{m,\sigma(m)}-Q_{m,\emptyset}) + \nabla_{K^m} u_m(K^m_{m,\sigma(m)}- K^m_{m,\emptyset}) 
+ 
\\ 
& + \nabla_{k^m} u_m (k^m_{m,\sigma(m)} - k^m_{m,\emptyset} - k^{m}_{m',\sigma(m)}) \end{split}
\end{equation*}
Then, individual rationality on one hand implies that 
$$
0 \le u_m(q^m_{m,\sigma(m)}, Q_{m,\sigma(m)}, k^m_{m,\sigma(m)}, K^m_{m,\sigma(m)}) -  u_m(q_{m,\emptyset}, Q_{m,\emptyset}, k^m_{m,\emptyset} + k^m_{m',\sigma(m)}, K^m_{m,\emptyset})
$$
At rewriting the right-hand side of the inequality we obtain,
\begin{align*}
0 & \le \lambda_{m,\emptyset} \big( p_{m,\emptyset} (q^m_{m,\sigma(m)}-q^m_{m,\emptyset}) + P_{m,\emptyset} (Q_{m,\sigma(m)}-Q_{m,\emptyset}) + \alpha_m (K^m_{m,\sigma(m)}- K^m_{m,\emptyset}) + \\ 
& \quad + (k^m_{m,\sigma(m)} - k^m_{m,\emptyset} - k^{m}_{m',\sigma(m)}) \big), \\
0 & \leq p_{m,\emptyset} (q^m_{m,\sigma(m)}-q^m_{m,\emptyset}) + P_{m,\emptyset} (Q_{m,\sigma(m)}-Q_{m,\emptyset}) + \rho^m_{m,\sigma(m)} (K^m_{m,\sigma(m)}- K^m_{m,\emptyset})
\\ & \quad + (k^m_{m,\sigma(m)} - k^m_{m,\emptyset} - k^{m}_{m',\sigma(m)} )
\end{align*}
Note that,
$$
y_{m,\emptyset} = p_{m,\emptyset} q^m_{m,\emptyset} + P_{m,\emptyset} Q_{m,\emptyset} + k^m_{m,\emptyset} + \rho^m_{m,\sigma(m)}  K^m_{m,\emptyset} \text { and } k^{m}_{m',\sigma(m)} \ge 0.
$$
Then we can rewrite the inequality as follows.
$$
y_{m,\emptyset} \le y_{m,\emptyset} + k^{m}_{m',\sigma(m)} \le p_{m,\emptyset} q^m_{m,\sigma(m)} + P_{m,\emptyset} Q_{m,\sigma(m)} + k^m_{m,\sigma(m)} + \rho^m_{m,\sigma(m)} K^m_{m,\sigma(m)}. 
$$

\paragraph{No Blocking Pairs.}
The optimization problem for a potential match $(m,w)\in M \times W$ looks as follows:

\begin{equation*}
\begin{cases}
& u_m(q^m,Q,k^m_m+k^m_{\sigma(m)},K^m)  +  \mu u_w(q^w,Q,k^w_w+k^w_{\sigma(w)},K^w) \rightarrow \max\limits_{q^m,q^w,Q,k^m_m,k^w_w,K^m,K^w} \\ 
& p_{m,w}(q^m + q^w) +P_{m,w}Q + k^m_m + k^w_w +\rho^m_{m,\sigma(m)} K^m + \rho^w_{\sigma(w),w} K^w \leq y_{m,w} \\
\end{cases}
\end{equation*}
Let ($q^m_{m,w}, q^w_{m,w}, Q_{m,w}, k^m_{m,w}, k^w_{m,w},K^m_{m,w}, K^w_{m,w}$) be the outcome of this optimization problem.
The first-order conditions look as follows.
\begin{equation*}
\begin{split}
&\nabla_{q^m} u_m(q^m_{m,w},Q_{m,w},k^m_{m,w} + k^m_{m',\sigma(m)},K^m_{m,w}) = \lambda_{m,w} p_{m,w}; \\
&\nabla_{q^w} \mu u_w(q^w_{m,w},Q_{m,w},k^w_{m,w}+k^w_{\sigma(w),w'},K^w_{m,w}) = \lambda_{m,w} p_{m,w}; \\
&\nabla_{Q} u_m(q^m_{m,w},Q_{m,w},k^m_{m,w} + k^m_{m',\sigma(m)},K^m_{m,w}) + \mu\nabla_{ Q} u_w(q^w_{m,w},Q_{m,w},k^w_{m,w}+k^w_{\sigma(w),w'},K^w_{m,w}) = \lambda_{m,w} P_{m,w}; \\
&\nabla_{k^m} u_m(q^m_{m,w},Q_{m,w},k^m_{m,w}+k^m_{m',\sigma(m)},K^m_{m,w}) = \lambda_{m,w};\\
&\nabla_{k^w} \mu u_w(q^w_{m,w},Q_{m,w},k^w_{m,w}+k^w_{\sigma(w),w'},K^w_{m,w}) = \lambda_{m,w}; \\
&\nabla_{K^m} u_m(q^m_{m,w},Q_{m,w},k^m_{m,w}+k^m_{m',\sigma(m)},K^m_{m,w}) = \lambda_{m,w} \rho^m_{m,\sigma(m)} ;\\
&\nabla_{K^w} \mu u_w(q^w_{m,w},Q_{m,w},k^w_{m,w}+k^w_{\sigma(w),w'},K^w_{m,w}) = \lambda_{m,w} \rho^w_{\sigma(w),w} ;
\end{split}
\end{equation*}
where $k^m_{m',\sigma(m)}$ is the contribution to own children's consumption by $\sigma(m)$ when matched with $m'$ and $k^w_{\sigma(w),w'}$ is the contribution to own children's consumption by $\sigma(w)$ when matched with $w'$.
Let 
$$P^w_{m,w} = \frac{\mu \nabla u_w(q^w_{m,w},Q_{m,w},k^w_{m,w} + k^w_{\sigma(w),w'}, K^w_{m,w})}{\lambda_{m,w}} \text{ and } P^m_{m,w} = P_{m,w} - P^w_{m,w}.$$
We know that in the marriage market setting if $(m,w)\in M\times W$ cannot be a blocking pair and the resulting Pareto frontier is continuous and strictly decreasing, then \cite{alkan1990core} implies that there is a $\mu>0$ such that the optimal consumption vector ($q^m_{m,w}, q^w_{m,w}, Q_{m,w}, k^m_{m,w}, k^w_{m,w},K^m_{m,w}, K^w_{m,w}$) satisfy
\begin{equation*}
\begin{cases}
& u_m(q^m_{m,w}, Q_{m,w}, k^m_{m,w} + k^{m}_{m',\sigma(m)},K^m_{m,w}) \le u_m(q^m_{m,\sigma(m)}, Q_{m,\sigma(m)}, k^m_{m,\sigma(m)}, K^m_{m,\sigma(m)}) \\    
& \mu u_w (q^w_{m,w},Q_{m,w},k^w_{m,w} + k^w_{\sigma(w),w'},K^w_{m,w}) \le 
\mu u_w (q^w_{\sigma(w),w}, Q_{\sigma(w),w}, k^w_{\sigma(w),w}, K^w_{\sigma(w),w})
\end{cases}
\end{equation*}
with at least one inequality being strict.
Note that $k^m_{m,\sigma(m)}$ is the total daily expenditures on $m$'s own children in the current household $(m,\sigma(m))$. Thus, $k^m_{m,\sigma(m)} = k^{\sigma(m)}_{m,\sigma(m)} = k_{m,\sigma(m)}$. Similarly, $k^w_{\sigma(w),w}$ is the total daily expenditures on $w$'s own children in the current household ($\sigma(w),w$). Thus, $k^w_{\sigma(w),w} = k^{\sigma(w)}_{\sigma(w),w} = k_{\sigma(w),w}$. The same logic holds for the big-decision expenditures.
Thus, given the concavity of the utility, first-order conditions and the implications of no blocking pairs we can infer that 
\begin{align*}
0 & \le 
u_m(q^m_{m,\sigma(m)}, Q_{m,\sigma(m)}, k^m_{m,\sigma(m)}, K^m_{m,\sigma(m)}) - 
u_m(q^m_{m,w}, Q_{m,w}, k^m_{m,w} + k^{m}_{m',\sigma(m)},K^m_{m,w}),
\\ & \le
p_{m,w}( q^m_{m,\sigma(m)} - q^m_{m,w}) + P^m_{m,w} ( Q_{m,\sigma(m)} - Q_{m,w}) + \rho^m_{m,\sigma(m)} ( K^m_{m,\sigma(m)} - K^m_{m,w}) \\
& \quad + (k^m_{m,\sigma(m)} - k^m_{m,w} - k^m_{m',\sigma(m)})
\end{align*}
and
\begin{align*}
0 & \le 
u_w (q^w_{\sigma(w),w}, Q_{\sigma(w),w}, k^w_{\sigma(w),w}, K^w_{\sigma(w),w}) - 
u_w (q^w_{m,w},Q_{m,w},k^w_{m,w} + k^w_{\sigma(w),w'},K^w_{m,w}), 
 \\ & \le
p_{m,w}( q^w_{\sigma(w),w} - q^w_{m,w}) + P^w_{m,w} ( Q_{\sigma(w),w} - Q_{m,w}) + \rho^w_{\sigma(w),w} ( K^w_{\sigma(w),w} - K^w_{m,w}) \\
& \quad + (k^w_{\sigma(w),w} - k^w_{m,w} - k^w_{\sigma(w),w'}).
\end{align*}
Adding up the inequalities and taking into account that $ k^m_{m',\sigma(m)}\ge 0$ and $k^w_{\sigma(w),w'} \ge 0$ we obtain
%
\begin{equation*}
	\begin{split}
y_{m, w} & \le y_{m,w} + k^m_{m',\sigma(m)} + k^w_{\sigma(w),w'}  \\
& \le p_{m,w} (q^m_{m,\sigma(m)} + q^w_{\sigma(w),w})  + P^m_{m,w} Q_{m,\sigma(m)} +  P^w_{m,w} Q_{\sigma(w),w} + \\
& \quad + k^m_{m,\sigma(m)} + \rho^m_{m,\sigma(m)} K^m_{m,\sigma(m)} + k^w_{\sigma(w),w} + \rho^w_{\sigma(w),w} K^w_{\sigma(w),w}
 	\end{split} 
\end{equation*}

\subsection{Proof of Proposition \ref{prop:SPC}}
Let $\tau_{m,w}^{m',\sigma(m)}$ be the child support payment from $m$ to $\sigma(m)$ when he forms a couple with $w$ and expects his ex-partner $\sigma(m)$ to form a couple with $m'$. 
By the assumption of perfect enforcement of custody law we have $\tau_{m,w}^{m',\sigma(m)} \geq T_{m,\sigma(m)}$, where $T_{m,\sigma(m)}$ is the minimum transfers a non-custodian parent is obliged to pay to a custodian parent. 
Denote by $D_{m',\sigma(m)}(\tau)$ the demand for children's consumption $\sigma(m)$ exhibits if she forms a couple with $m'$ ($m' \neq m$) and receives a transfer $\tau$ from $m$. Let $D^{-1}_{m',\sigma(m)}(C)$ be the inverse of this demand function.
Moreover, we assume $D_{m',\sigma(m)}$ to be smooth for all $(m',\sigma(m))$, hence the inverse is smooth as well.
Furthermore, we assume that $D_{m',\sigma(m)}$ is a concave function and $D'_{m',\sigma(m)}(\tau)\geq 1$. 
This assumption essentially means that $\sigma(m)$ is compliant. 
That is, she does not use the money designated for children's consumption for her personal consumption.
This can be guaranteed, for instance, by the presence of an alternate possibility for $m$ to buy the children's consumption at the market price himself.

\paragraph{Individual Rationality for $m\in M$.}
$m$ gets utility from the consumption of private and public goods and children's consumption (which is produced through cash transfers to his ex-partner $\sigma(m)$). We assume that children's consumption is a single-dimensional Hicksian good with price normalized to one.
Although children's consumption cannot be brought directly from the market, it can be controlled by $m$ through a transfer to $\sigma(m)$.
We assume that the consumption decisions, upon divorce, can be represented as the subgame-perfect equilibrium of a two-stage game.
In the first stage, $m$ optimizes his utility function conditional on the best-response taken by $\sigma(m)$ for every cash transfer.
In the second stage, $\sigma(m)$ optimizes her utility function given the transfer by $m$.
The optimization problem of $m$, when single, looks like:

\begin{equation*}
\begin{cases}
&  u_m(q^m,Q,C^m) \rightarrow \max_{q^m,Q,\tau} \\ 
& p_{m,\emptyset}q^m +P_{m,\emptyset}Q + \tau \leq y_{m,\emptyset} \\
&C^m = D_{m',\sigma(m)}(\tau) \\
&\tau \geq T_{m,\sigma(m)}
\end{cases}
\end{equation*}
Let $(q^m_{m,\emptyset}, Q_{m,\emptyset}, \tau^{m', \sigma(m)}_{m,\emptyset})$ be the solution to this problem. First order conditions give,

\begin{equation*}
\begin{split}
&\nabla_{q^m} u_m(q^m_{m,\emptyset},Q_{m,\emptyset},C^m_{m,\emptyset}) = \lambda^1_{m,\emptyset} p_{m,\emptyset} \\
&\nabla_{Q} u_m(q^m_{m,\emptyset},Q_{m,\emptyset},C^m_{m,\emptyset}) = \lambda^1_{m,\emptyset} P_{m,\emptyset} \\
&\nabla_{C^m} u_m(q^m_{m,\emptyset},Q_{m,\emptyset},C^m_{m,\emptyset}) = \frac{(\lambda^1_{m,\emptyset}-\lambda^2_{m,\emptyset})}{ D'_{m',\sigma(m)}(\tau_{m,\emptyset}^{m',\sigma(m)})}
\end{split}
\end{equation*}
where $\lambda^1_{m,\emptyset} \geq 0$ and $\lambda^2_{m,\emptyset} \geq 0$ are Lagrange multipliers for the budget constraint and the minimum transfer constraint.
Let  $\varphi_{m,\emptyset}^{m',\sigma(m)} = \frac{1}{D'_{m',\sigma(m)}(\tau_{m,\emptyset}^{m',\sigma(m)})}$.
Since $1 \leq D'_{m',\sigma(m)} $, we have, $0< \varphi_{m,\emptyset}^{m',\sigma(m)} \leq 1$.
We further assume that the minimum transfer constraint $\tau \geq T_{m,\sigma(m)}$ is not binding ($\lambda^2_{m,\emptyset} = 0$).\footnote{ 
Later, we will present the conditions without this assumption. We show that with the binding minimum transfers constraint, the resulting conditions are weaker in nature. Thus, using those conditions for identification would gives less informative bounds.}
Concavity of the utility function implies, 
\begin{equation*}
\begin{split}
& u_m(q^m_{m,\sigma(m)}, Q_{m,\sigma(m)}, C_{m,\sigma(m)}) -  u_m(q_{m,\emptyset}, Q_{m,\emptyset}, C_{m,\emptyset})  \\ & \leq
\lambda^1_{m,\emptyset} p_{m,\emptyset} (q^m_{m,\sigma(m)}-q_{m,\emptyset}) + \lambda^1_{m,\emptyset} P_{m,\emptyset} (Q_{m,\sigma(m)}-Q_{m,\emptyset}) + \lambda^1_{m,\emptyset} \varphi^{m',\sigma(m)}_{m,\emptyset} (C_{m,\sigma(m)}-C^m_{m,\emptyset}) \\
\end{split}
\end{equation*}
By individual rationality of $m$,
\begin{equation*}
\begin{split}
& 0 \leq u_m(q^m_{m,\sigma(m)}, Q_{m,\sigma(m)}, C_{m,\sigma(m)}) -  u_m(q_{m,\emptyset}, Q_{m,\emptyset}, C^m_{m,\emptyset})  \\ 
& 0 \leq
p_{m,\emptyset} (q^m_{m,\sigma(m)}-q_{m,\emptyset}) + P_{m,\emptyset} (Q_{m,\sigma(m)}-Q_{m,\emptyset}) + \varphi^{m',\sigma(m)}_{m,\emptyset}  (C_{m,\sigma(m)}-C^m_{m,\emptyset}) \\
& p_{m,\emptyset} q_{m,\emptyset} + P_{m,\emptyset}Q_{m,\emptyset}  \leq
p_{m,\emptyset}q^m_{m,\sigma(m)} + P_{m,\emptyset}Q_{m,\sigma(m)} +\varphi^{m',\sigma(m)}_{m,\emptyset}  (C_{m,\sigma(m)}-C^m_{m,\emptyset}) \\
&y_{m,\emptyset}-\tau_{m,\emptyset}^{m',\sigma(m)} \leq p_{m,\emptyset}q^m_{m,\sigma(m)}+P_{m,\emptyset}Q_{m,\sigma(m)} + \varphi_{m,\emptyset}^{m',\sigma(m)} (C^m_{m,\sigma(m)}-C^m_{m,\emptyset})
\end{split}
\end{equation*}
In order to simplify the last term on the right hand of the inequality, first 
recall that $0 < \varphi_{m,\emptyset}^{m',\sigma(m)}\leq 1$. 
Thus, we can write $\varphi_{m,\emptyset}^{m',\sigma(m)} C^m_{m,\sigma(m)}\leq C^m_{m,\sigma(m)}$.
Next,
$$
\tau = D_{m',\sigma(m)}^{-1}(C) = D_{m',\sigma(m)}^{-1}(0) + \int_{0}^{\tau} \frac{1}{D'_{m',\sigma(m)}(x)} dx \leq \frac{C}{D'_{m',\sigma(m)}(\tau)}
$$
Here, we make a technical assumption that $D^{-1}(0)=0$. 
The latter inequality is implied by concavity of $D_{m,w}$ (that is, $D'_{m,w}(\tau) \leq D'_{m,w}(\tau')$ for every $\tau'<\tau$.)
This gives,
\begin{equation*}
\begin{split}
& \tau_{m,\emptyset}^{m',\sigma(m)} = D_{m',\sigma(m)}^{-1}(C^m_{m,\emptyset}) \leq \frac{C^m_{m,\emptyset}}{D'_{m',\sigma(m)}(\tau_{m,\emptyset}^{m',\sigma(m)})} \\
& \tau_{m,\emptyset}^{m',\sigma(m)} \leq \varphi_{m,\emptyset}^{m',\sigma(m)}{C^m_{m,\emptyset}} 
\end{split}
\end{equation*}
Substituting these inequalities and further simplification gives the individual rationality condition shown in Proposition \ref{prop:SPC}. 
\begin{equation*}
\begin{split}
& y_{m,\emptyset}-\tau_{m,\emptyset}^{m',\sigma(m)}  \leq p_{m,\emptyset}q^m_{m,\sigma(m)}+P_{m,\emptyset}Q_{m,\sigma(m)} + C^m_{m,\sigma(m)}-\tau_{m,\emptyset}^{m',\sigma(m)} \\
& y_{m,\emptyset} \leq p_{m,\emptyset}q^m_{m,\sigma(m)}+P_{m,\emptyset}Q_{m,\sigma(m)} + C^m_{m,\sigma(m)}
\end{split}
\end{equation*}

\paragraph{Individual Rationality for $w\in W$.}
Recall that the consumption decisions of $w\in W$ are made in the second stage when she receives the unconditional income transfer ($\tau_{\sigma(w),w'}^{\emptyset,w}$) from $\sigma(w)$.
The optimization problem of $w$, when single, looks as follows:
\begin{equation*}
\begin{cases}
&  u_w(q^w,Q,C^w) \rightarrow \max_{q^w,Q,C^w} \\ 
& p_{\emptyset,w}q^w +P_{\emptyset,w}Q + C^w \leq y_{\emptyset,w} + \tau_{\sigma(w),w'}^{\emptyset,w} 
\end{cases}
\end{equation*}
Let $(q^w_{\emptyset,w}, Q_{\emptyset,w}, C^w_{\emptyset,w})$ be the solution to the above problem. First order conditions imply,
\begin{equation*}
\begin{split}
&\nabla_{q^w} u_w(q^w_{\emptyset,w},Q_{\emptyset,w},C^w_{\emptyset,w}) = \lambda_{\emptyset,w} p_{\emptyset,w} \\
&\nabla_{Q} u_w(q^w_{\emptyset,w},Q_{\emptyset,w},C^w_{\emptyset,w}) = \lambda_{\emptyset,w} P_{m,\emptyset} \\
&\nabla_{C} u_w(q^w_{\emptyset,w},Q_{\emptyset,w},C^w_{\emptyset,w}) = \lambda_{\emptyset,w} 
\end{split}
\end{equation*}
Using concavity of $u_w$ and individual rationality condition for $w$ (in a similar way as used before), we get the following inequality: 
\begin{equation*}
y_{\emptyset,w} + \tau_{\sigma(w),w'}^{\emptyset,w} \leq p_{\emptyset,w}q^w_{\sigma(w),w}+P_{\emptyset,w}Q_{\sigma(w),w} + C_{\sigma(w),w}
\end{equation*}
Finally, using the assumption of perfect enforcement of law ($\tau_{\sigma(w),w'}^{\emptyset,w}\geq T_{\sigma(w),w}$), we get the individual rationality conditions for $w$ as shown in Proposition \ref{prop:SPC}.\footnote{
Note that we can also use $\tau_{\sigma(w),w'}^{\emptyset,w}$ as unknown transfers (with minimum binding constraint) and still get linear conditions. 
However, this would make no difference from the computational point of view, because the algorithms to solve the linear inequalities would equalize $\tau_{\sigma(w),w'}^{\emptyset,w}$ to $T_{\sigma(w),w}$ in order to maximize the stability.
Moreover, the unknown transfer $\tau_{\sigma(w),w'}^{\emptyset,w}$ depends on the belief about the potential match ($w'$) of $\sigma(w)$. 
This requires us to make significant additional assumptions in order to apply conditions without adding anything to the empirical content.
}
\begin{equation*}
y_{\emptyset,w} + T_{\sigma(w),w} \leq p_{\emptyset,w}q^w_{\sigma(w),w}+P_{\emptyset,w}Q_{\sigma(w),w} + C_{\sigma(w),w}
\end{equation*}
%
%

\paragraph{No Blocking Pairs.} 
The optimization problem of a potential match $(m,w)$ looks as follows:
\begin{equation*}
\begin{cases}
& u_m(q^m, Q, C^m) + \mu u_w(q^w, Q, C^w) \rightarrow \max_{q^m, q^w, Q, \tau, C^w}  \\ 
& p_{m,w}(q^m+q^w) +P_{m,w}Q + \tau + C^w\leq y_{m,w} + \tau_{\sigma(w),w'}^{m,w} \\
& C^m = D_{m',\sigma(m)}(\tau) \\
&  \tau \geq T_{m,\sigma(m)} \\
\end{cases}
\end{equation*}
Here $m'$ is the belief of $m$ about potential rematch of his ex-partner $\sigma(m)$. Similarly, $w'$ indicates the potential rematch of $w$'s ex-partner $\sigma(w)$. The first order condition would look as follows.
\begin{equation*}
\begin{split}
&\nabla_{q^m} u_m(q^m_{m,w},Q_{m,w},C^m_{m,w}) = \lambda^1_{m,w} p_{m,w}; \\
&\nabla_{q^w} \mu u_w(q^w_{m,w},Q_{m,w},C^{w}_{m,w}) = \lambda^1_{m,w} p_{m,w}; \\
&\nabla_{Q} u_m(q^m_{m,w},Q_{m,w},C^m_{m,w}) + \mu\nabla_{ Q} u_w(q^w_{m,w},Q_{m,w},C^{w}_{m,w}) = \lambda^1_{m,w} P_{m,w}; \\
&\nabla_{C^m} u_m(q^m_{m,w},Q_{m,w},C^m_{m,w}) = \frac{\lambda^1_{m,w} - \lambda^2_{m,w}}{D'_{m',\sigma(m)}(\tau)}\\
&\nabla_{C^w} \mu u_w(q^w_{m,w},Q_{m,w},C^{w}_{m,w}) = \lambda^1_{m,w};
\end{split}
\end{equation*}
Denote $\varphi_{m,w}^{m',\sigma(m)} = \frac{1}{D'_{m',\sigma(m)}(\tau_{m,w}^{m',\sigma(m)})}$.
Since $D'_{m',\sigma(m)}(\tau_{m,w}^{m',\sigma(m)}) \geq 1$, we have $0 < \varphi_{m,w}^{m',\sigma(m)} \leq 1$.
Let $P^w_{m,w} = \frac{\mu \nabla u_w(q^w_{m,w},Q_{m,w},C^w_{m,w})  }{\lambda_{m,w}}$ and $P^m_{m,w} = P_{m,w} - P^w_{m,w}$.
Finally, since the minimum transfers constraint is assumed to be non-binding, we have $\lambda^2_{m,w} = 0$.
Using concavity of utility functions and no blocking pair restrictions we can obtain the following inequalities using the same logic as for the proof of Proposition \ref{prop:JC}.
\begin{equation*}
\begin{split}
y_{m,w}- \tau_{m,w}^{m',\sigma(m)}+ \tau_{\sigma(w),w'}^{m,w} &\leq p_{m,w}(q^m_{m,\sigma(m)} + q^w_{\sigma(w),w}) +P^m_{m,w}Q_{m,\sigma(m)} +P^m_{m,w}Q_{\sigma(w),w}  + \\ &+ C_{\sigma(w),w} +  \varphi_{m,w}^{m',\sigma(m)}(C_{m,\sigma(m)}-C^m_{m,w})
\end{split}
\end{equation*}
Using similar arguments as in individual rationality conditions for $m$, we have
\begin{equation*}
\begin{split}
 & \varphi_{m,w}^{m',\sigma(m)}C^m_{m,w} \geq \tau_{m,w}^{m',\sigma(m)} \\ 
 & \varphi_{m,w}^{m',\sigma(m)}C_{m,\sigma(m)} \leq C_{m,\sigma(m)} \\ 
 & \tau_{\sigma(w),w'}^{m,w} \geq T_{\sigma(w),w} 
 \end{split}
\end{equation*} 
Using these inequalities, we get the no blocking pairs condition shown in Proposition \ref{prop:SPC}.
\begin{equation*}
\begin{split}
y_{m,w} + T_{\sigma(w),w} \leq p_{m,w}(q^m_{m,\sigma(m)} + q^w_{\sigma(w),w})+P^m_{m,w}Q_{m,\sigma(m)} +P^m_{m,w}Q_{\sigma(w),w}  + C_{\sigma(w),w} + C_{m,\sigma(m)}
\end{split}
\end{equation*}

\paragraph{Binding Minimal Transfers.}
Now, we present the conditions when we assume that the minimum transfer constraints are binding.
In this case $\lambda^2_{m,w}>0$ and $\tau_{m,w}^{m',\sigma(m)} = T_{m,\sigma(m)}$.
Concavity of $u_m$ and individual rationality of $m$ implies,
\begin{equation*}
\begin{split}
y_{m,\emptyset}-T_{m,\sigma(m)} 
\leq p_{m,\emptyset}q^m_{m,\sigma(m)}+P_{m,\emptyset}Q_{m,\sigma(m)} + \frac{1- \frac{\lambda^2_{m,\emptyset}}{\lambda^1_{m,\emptyset}} }{ D_{m',\sigma(m)}(\tau^{m,\emptyset}_{m',\sigma(m)}))} \big[  C^m_{m,\sigma(m)} -  C_{m,\emptyset} \big]
\end{split}
\end{equation*}
Note that, $ \nabla_{C^m} u_m(q^m_{m,\emptyset}, Q_{m,\emptyset}, C^m_{m,\emptyset})) \leq \frac{\lambda^1_{m,w} - \lambda^2_{m,w}}{D'_{m',w}(\tau))}$. 
Since $u_m$ is strictly increasing in every component and $D'_{m',w}(\tau) \geq 1$, it must be the case that $\lambda^2_{m,w}\leq \lambda^1_{m,w}$.
Hence, $0 \leq \frac{1- \frac{\lambda^2_{m,w}}{\lambda^1_{m,w}} }{ D_{m',w}(\tau))}  \leq 1$. Simplifying above inequality gives, 
\begin{equation*}
y_{m,\emptyset}-T_{m,\sigma(m)} \leq p_{m,\emptyset}q^m_{m,\sigma(m)}+P_{m,\emptyset}Q_{m,\sigma(m)} + C_{m,\sigma(m)}
\end{equation*}
Individual rationality condition for $w\in W$ is the same as for the case of non-binding constraints on minimum transfers.
Using the same arguments given above, we can derive the no blocking pairs conditions in case minimum transfers are binding.
\begin{equation*}
\begin{split}
& y_{m,w}-T_{m,\sigma(m)} + T_{\sigma(w),w} \leq p_{m,w}(q^m_{m,\sigma(m)} + q^w_{\sigma(w),w} )+P^m_{m,w} Q_{m,\sigma(m)} +  P^w_{m,w} Q_{\sigma(w),w} \\
& +  C_{m,\sigma(m)} + C_{\sigma(w),w}
\end{split}
\end{equation*}

\section{Empirical Illustrations - Additional results}
\label{appendix_empirics}
\subsection{Data and Empirical Setup}
\paragraph{LISS.} Table \ref{table_sumstat_LISS} provides summary statistics for all the selected couples from the LISS panel. Wages are net hourly wages. Leisure is measured in hours per week. To compute leisure hours, we assume that an individual needs 8 hours per day for sleeping and personal care (i.e., leisure = 168 - 56 hours worked). Non-labor income and (Hicksian) expenditures are measured in euros per week.
\begin{table}[htbp]\centering

\caption{Summary statistics for couples - LISS \label{table_sumstat_LISS}}
\begin{tabular}{l*{1}{cccc}}
\\ \toprule
                    &        mean&          sd&         min&         max\\
\midrule
male hourly wage    &       13.65&        4.32&        5.77&       36.06\\
female hourly wage  &       12.05&        3.30&        1.80&       26.96\\
male labor hours    &       40.61&       10.10&       10.00&       80.00\\
female labor hours  &       28.08&       10.73&       10.00&       73.00\\
male leisure hours  &       71.39&       10.10&       32.00&      102.00\\
female leisure hours&       83.92&       10.73&       39.00&      102.00\\
male age            &       46.60&        8.68&       26.00&       65.00\\
female age          &       44.44&        8.77&       26.00&       62.00\\
presence of children (1 = yes/ 0 = no)&        0.66&        0.47&        0.00&        1.00\\
number of children  &        1.31&        1.12&        0.00&        4.00\\
household expenditure&      785.35&      261.98&      236.54&     1910.77\\
children expenditure&      145.21&      136.34&        0.00&      684.79\\
non-labor income    &     -110.32&      342.16&    -1468.85&     1087.42\\
\bottomrule
\end{tabular}
\end{table}

\paragraph{RLMS.} Table \ref{table_sumstat_RLMS} provides summary statistics for all the selected couples from the LISS panel.
Wages are hourly wage rates in Russian ruble (average exchange rate for 2013 was 1 ruble for .0341 USD). Leisure is the average hours spent on leisure per week (computed as above). Non-labor income, household and children expenditure are measured in rubles per week. Further, the table also reports on age, education, presence, and number of children.  
\begin{table}[htbp]\centering

\caption{Summary statistics for couples - RLMS \label{table_sumstat_RLMS}}
\begin{tabular}{l*{1}{cccc}}
\\ \toprule
                    &        mean&          sd&         min&         max\\
\midrule
male hourly wage    &      129.58&       74.04&       23.81&      561.80\\
female hourly wage  &       98.16&       61.33&       17.39&      400.00\\
male labor hours    &       43.75&       11.80&       11.09&      110.85\\
female labor hours  &       39.89&       10.29&       10.39&       93.53\\
male leisure hours  &       68.25&       11.80&        1.15&      100.91\\
female leisure hours&       72.11&       10.29&       18.47&      101.61\\
male age            &       41.95&        9.40&       25.00&       65.00\\
female age          &       40.06&        9.47&       25.00&       65.00\\
male has a degree (1 = yes/ 0 = no)&        0.26&        0.44&        0.00&        1.00\\
female has a degree (1 = yes/ 0 = no)&        0.39&        0.49&        0.00&        1.00\\
presence of children (1 = yes/ 0 = no)&        0.66&        0.47&        0.00&        1.00\\
number of children  &        0.91&        0.78&        0.00&        3.00\\
household expenditure&    15868.89&    13675.58&      461.89&    93479.58\\
children expenditure&     2331.38&     3218.22&        0.00&    24549.05\\
adult expenditure   &    13537.51&    11652.98&      461.89&    85192.63\\
non-labor income    &     6630.79&    12121.75&    -7472.29&    80739.19\\
\bottomrule
\end{tabular}
\end{table}

\begin{figure}[htbp]
    \centering
    \caption{Region-based marriage markets}
    \label{figure_russia}
    \includegraphics[scale=0.3]{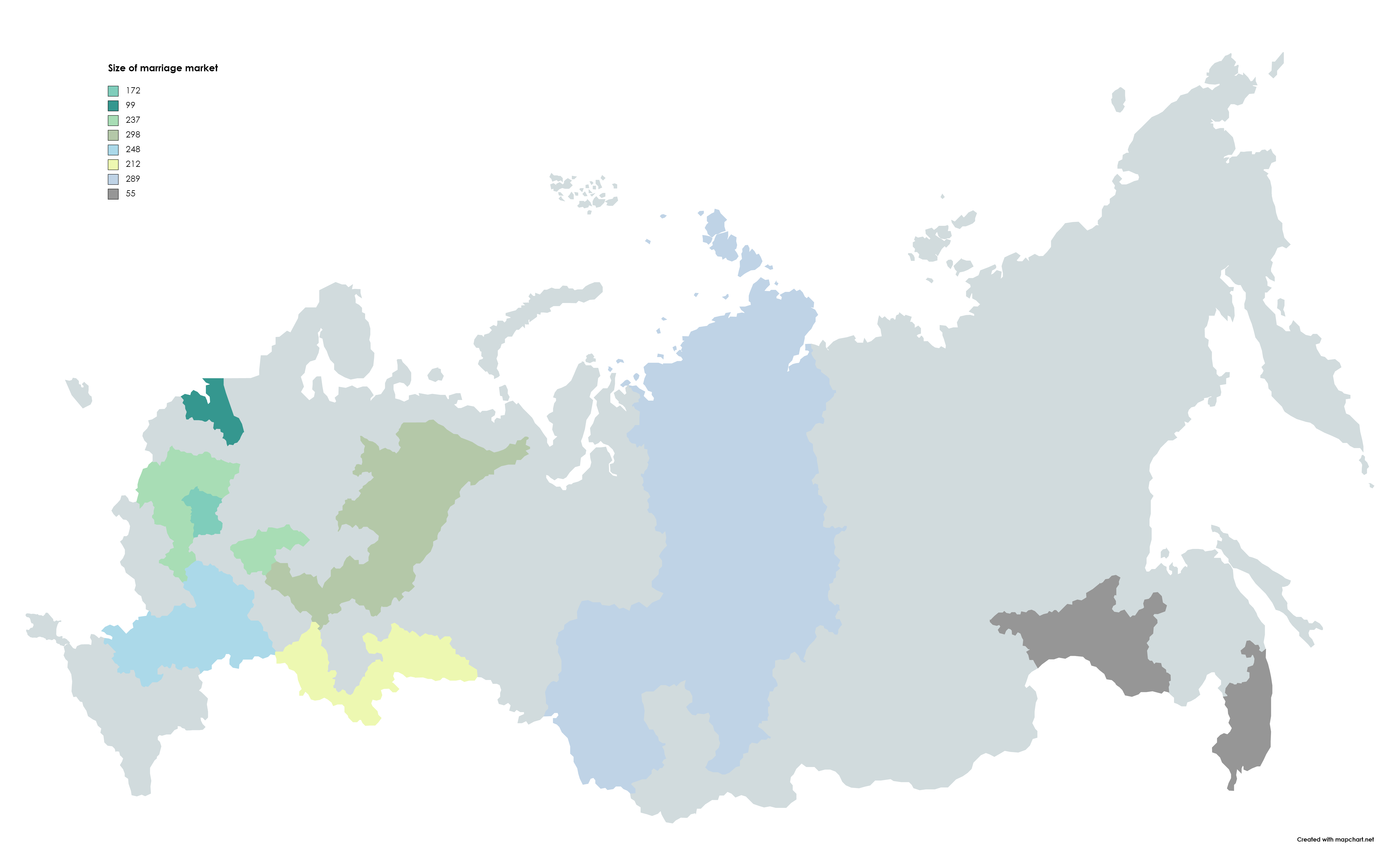}
\end{figure}

\paragraph{Marriage Markets.} Figure \ref{figure_russia} provides a map of the eight marriage markets considered and gives each market's subsample size. Table \ref{table_market_RLMS} details the regions included in the marriage market shown in Figure \ref{figure_russia}. It also reports the average sizes of males' and females' consideration sets within each market.
\begin{table}[htbp]\centering
\caption{Marriage markets - RLMS \label{table_market_RLMS}}
\begin{tabular}{p{9.5cm}cc} \\ \toprule marriage market & \multicolumn{2}{c}{average consideration sets} \\

     &        male&      female\\
\midrule
Moscow City,  Moscow Oblast&       45.61&       37.09\\
Leningrad Oblast: Volosovkij Rajon, St. Petersburg City&       27.88&       16.09\\
Kaluzhskaya Oblast: Kuibyshev Rajon, Gorkovskaja Oblast: Nizhnij Novgorod, Tulskaja Oblast: Tula, Kalinin Oblast: Rzhev C, Lipetskaya Oblast: Lipetsk CR, Smolensk CR&       75.30&       64.44\\
Perm Oblast: Solikamsk City and Rajon, Tatarskaja ASSR: Kazan, Chuvashskaya ASSR: Shumerlja CR, Komi-ASSR: Syktyvkar, Komi-ASSR: Usinsk CR, Udmurt ASSR: Glasov CR&      102.75&       74.81\\
Tambov Oblast: Uvarovo CR, Volgograd Oblast: Rudnjanskij Rajon, Saratov Oblast: Volskij Gorosovet and Rajon, Penzenskaya Oblast: Zemetchinskij Rajon, Rostov Oblast: Batajsk,  Saratov CR&       80.49&       62.54\\
Kurgan, Orenburg Oblast: Orsk, Cheliabinsk, Cheliabinsk Oblast:  Krasnoarmeiskij Rajon&       64.80&       39.34\\
Krasnojarskij Kraij: Krasnojarsk,  Altaiskij Kraj: Kurinskij Rajon, Krasnojarskij Kraij: Nazarovo CR, Altaiskij Kraj: Biisk CR, Tomsk, Berdsk City and Raion: Novosibirskaya Oblast&       90.70&       64.84\\
Vladivostok, Amurskaja Oblast: Tambovskii Rajon&       18.39&       16.13\\
Total&       76.77&       57.37\\
\bottomrule
\end{tabular}
\end{table}

\subsection{Female Sharing Rule and Observed Characteristics}

Tables \ref{table_bounds_shares} and \ref{table_bounds_qf} reveal quite some heterogeneity in the identified female allocation across households. We investigate this further by relating the lower and upper bounds of female sharing rule and private consumption share to the observed household characteristics. Tables \ref{table_regression_qf_LISS} and \ref{table_regression_qf_RLMS} report the regression estimates. The columns `lower' (`upper') correspond to the estimates obtained when we use the recovered lower (upper) bound as the dependent variable. 
\begin{table}[htbp]\centering

\caption{Females' private consumption, sharing rule and observed heterogeneity - LISS \label{table_regression_qf_LISS}}
\begin{tabular}{l*{4}{c}}
\toprule & \multicolumn{2}{c}{$ q^f/q $} & \multicolumn{2}{c}{$\eta^f $} \\ & lower & upper & lower & upper \\
\midrule
log$(w_f/w_m)$          &       0.140***&       0.228***&       0.186***&       0.185***\\
                    &     (0.031)         &     (0.033)         &     (0.008)         &     (0.011)         \\
\addlinespace
age male - age female&       0.001         &       0.003         &      -0.000         &      -0.001         \\
                    &     (0.004)         &     (0.004)         &     (0.001)         &     (0.001)         \\
\addlinespace
log(total potential income)&      -0.067         &      -0.017         &      -0.021         &      -0.017         \\
                    &     (0.041)         &     (0.042)         &     (0.015)         &     (0.015)         \\
\addlinespace
one child          &      -0.060*  &       0.006         &      -0.012*  &       0.029***\\
                    &     (0.031)         &     (0.033)         &     (0.007)         &     (0.007)         \\
\addlinespace
two or more children          &      -0.109***&      -0.017         &      -0.011*  &       0.047***\\
                    &     (0.027)         &     (0.027)         &     (0.006)         &     (0.005)         \\
\addlinespace
male's consideration set size&      -0.001         &      -0.001         &       0.000         &      -0.000         \\
                    &     (0.001)         &     (0.001)         &     (0.000)         &     (0.000)         \\
\addlinespace
females' consideration set size&       0.001** &       0.002***&      -0.000*  &       0.000         \\
                    &     (0.001)         &     (0.001)         &     (0.000)         &     (0.000)         \\
\addlinespace
constant            &       0.894***&       0.575*  &       0.607***&       0.714***\\
                    &     (0.329)         &     (0.333)         &     (0.118)         &     (0.120)         \\
\midrule
N                   &         239         &         239         &         239         &         239         \\
R-squared                  &       0.195         &       0.240         &       0.778         &       0.816         \\
\bottomrule
\multicolumn{5}{l}{\footnotesize Standard errors in parentheses}\\
\multicolumn{5}{l}{\footnotesize * \(p<0.10\), ** \(p<0.05\), *** \(p<0.01\)}\\
\end{tabular}
\end{table}

\begin{table}[htbp]\centering

\caption{Females' private consumption, sharing rule and observed heterogeneity - RLMS \label{table_regression_qf_RLMS}}
\begin{tabular}{l*{4}{c}}
\toprule & \multicolumn{2}{c}{$ q^f/q $} & \multicolumn{2}{c}{$\eta^f $} \\ & lower & upper & lower & upper \\
\midrule
log$(w_f/w_m)$          &       0.149***&       0.215***&       0.153***&       0.173***\\
                    &     (0.011)         &     (0.012)         &     (0.005)         &     (0.005)         \\
\addlinespace
age male - age female&      -0.002*  &      -0.000         &      -0.000         &       0.000         \\
                    &     (0.001)         &     (0.002)         &     (0.001)         &     (0.001)         \\
\addlinespace
log(total potential income)&       0.023*  &       0.183***&      -0.051***&       0.084***\\
                    &     (0.012)         &     (0.016)         &     (0.006)         &     (0.006)         \\
\addlinespace
one child          &       0.079***&       0.104***&       0.013** &       0.052***\\
                    &     (0.013)         &     (0.016)         &     (0.006)         &     (0.005)         \\
\addlinespace
two or more children          &       0.096***&       0.182***&       0.019** &       0.084***\\
                    &     (0.017)         &     (0.020)         &     (0.008)         &     (0.007)         \\
\addlinespace
male has a degree          &      -0.014         &      -0.024         &       0.010         &      -0.006         \\
                    &     (0.013)         &     (0.017)         &     (0.006)         &     (0.006)         \\
\addlinespace
female has a degree          &       0.030** &       0.015         &       0.020***&      -0.014***\\
                    &     (0.012)         &     (0.015)         &     (0.005)         &     (0.005)         \\
\addlinespace
male's consideration set size&      -0.002***&      -0.003***&      -0.000         &      -0.001***\\
                    &     (0.000)         &     (0.000)         &     (0.000)         &     (0.000)         \\
\addlinespace
females' consideration set size&       0.002***&       0.002***&       0.001***&       0.000         \\
                    &     (0.000)         &     (0.001)         &     (0.000)         &     (0.000)         \\
\addlinespace
cohabitating  &       0.032*  &       0.019         &       0.015** &      -0.005         \\
                    &     (0.016)         &     (0.021)         &     (0.008)         &     (0.007)         \\
\addlinespace
constant            &      -0.036         &      -1.549***&       0.847***&      -0.295***\\
                    &     (0.122)         &     (0.173)         &     (0.059)         &     (0.065)         \\
\addlinespace
market fixed effect &         Yes         &         Yes         &         Yes         &         Yes         \\
\midrule
N                   &         879         &         879         &         879         &         879         \\
R-squared                  &       0.368         &       0.481         &       0.669         &       0.739         \\
\bottomrule
\multicolumn{5}{l}{\footnotesize Standard errors in parentheses}\\
\multicolumn{5}{l}{\footnotesize * \(p<0.10\), ** \(p<0.05\), *** \(p<0.01\)}\\
\end{tabular}
\end{table}

\end{document}